\pdfoutput=1
\documentclass[journal]{IEEEtran}

\usepackage[normalem]{ulem}
\usepackage{color}
\usepackage{graphics} 
\usepackage{epsfig} 
\usepackage{mathptmx} 
\usepackage{times} 
\usepackage{amsmath} 
\usepackage{amssymb}  
\usepackage{algorithm2e}
\usepackage{bbm}
\usepackage{multicol}

\usepackage{cite}
\usepackage{algorithmic}

\usepackage{array}

\usepackage[tight,footnotesize]{subfigure}
\begin{document}
%
\title{ Optimal Provision of {\color{black}Regulation Service} Reserves Under Dynamic Energy Service Preferences}
%
%
%

\author{Bowen~Zhang,~\IEEEmembership{Member,~IEEE,}
        Michael~C.~Caramanis,~\IEEEmembership{Senior~Member,~IEEE,}
        and~John~Baillieul,~\IEEEmembership{Life Fellow,~IEEE}
\thanks{B. Zhang is a visiting scholar at Department of Mechanical Engineering, Boston University, Boston, MA, 02215 USA e-mail: bowenz@bu.edu.}

\thanks{M. C. Caramanis is with the Department of Mechanical Engineering and the Division
of Systems Engineering, Boston University, Boston,
MA, 02215 USA e-mail: mcaraman@bu.edu.}

\thanks{J. Baillieul is with the Department of Electrical and Computer Engineering, the Department of Mechanical Engineering, and the Division
of Systems Engineering, Boston University, Boston,
MA, 02215 USA e-mail: johnb@bu.edu.}

\thanks{Research reported here was supported by NSF grant 1038230.}

\thanks{Manuscript received ********; revised ********.}}

\maketitle

\begin{abstract}
\color{black}We propose and solve a stochastic dynamic programming (DP) problem addressing the optimal provision of bidirectional regulation service reserves (RSR) by pools of duty cycle appliances.  Using our previously introduced concept of packetized energy, the approach (i) models the dynamics of the utility $U(T)$ that each consumer associates with a block/packet of energy when the prevailing temperature is $T$ in the appliance specific comfort zone, and (ii) derives a dynamic pricing policy that closes the loop between time varying appliance specific utilities and a regulation signal broadcast by the Independent System Operator (ISO) every 4 seconds. Since $T$ evolves dynamically as a function of past energy block purchases and heat losses, $T$ and $U(T)$ are time varying. Consumer utility levels $U(T)$ are modeled by the probability distribution of utility levels across the ensemble of all idle appliances at time $t$. This probability distribution provides a statistic of the detailed appliance specific utility information which is sufficient for the determination of an optimal pricing policy. Moreover, it can capture appliance specific diversity within comfort zone limits and temperature-utility function relationships. Dynamic pricing policies investigated in the literature assume that this ensemble probability distribution is time invariant and in fact uniform. A contribution of the work that follows is the replacement of this unrealistic simplifying assumption with the introduction of a new state variable modeling the aforementioned dynamic probability density function. The new state variable evolves as a function of the dynamic pricing policy which aims to track the centrally broadcast regulation signal trading off optimally between tracking error and appliance energy consumption utility. We specifically model active and idle appliances by means of a closed queuing system with price-controlled transitions from the idle to the active queue. Empirical evidence indicates that the dynamically changing frequency distribution of appliance specific utilities is closely approximated by a trapezoid-shape distribution function characterized by a single dynamic parameter which is in fact the aforementioned sufficient statistic. The optimal provision of RSR is addressed within a DP framework where we (i) derive an analytic characterization of the optimal pricing policy and the differential cost function, and (ii) prove optimal policy monotonicity and value function convexity. These properties enable us to implement a smart assisted value iteration (AVI) and an approximate DP (ADP) algorithm that exploit related functional approximations. Numerical results demonstrate (1) the validity of the solution techniques and the computational advantage of the proposed ADP on realistic, large-state-space problems, and (2) that the RSR provision based on the optimal control that respects dynamic consumer preferences has higher tracking accuracy relative to approaches assuming static uniformly distributed  consumer preferences.
\end{abstract}


%
\IEEEpeerreviewmaketitle

\section{Introduction}
The urgently needed reduction of $\textrm{CO}_{2}$ emissions will rely on the adoption of significant renewable electricity generation, whose volatility and intermittency will result in commensurate increase in the provision of secondary or Regulation Service Reserve (RSR) required to ensure power system stability through adequate Frequency Control (FC) and Area Control Error (ACE) management \cite{MakarovLoutan2009}, \cite{6344606}, \cite{5713846}, \cite{hammerstrom2007pacific}. The provision of these additional RSRs from centralized generation units, until now has been the primary approach, is expensive and poses a major challenge to massive renewable integration. The option of employing much lower cost demand-control-based RSR is rightly a major Federal Energy Regulatory Commission target \cite{FERC_2013}. Not surprisingly, there is extensive literature on demand management addressing, for example, direct load control (DLC) of thermostatic appliances to provide frequency reserve or load shedding \cite{LiCaramanis2012}, \cite{BilginCaramanis2012}, \cite{BowenBaillieul2012}, \cite{Chassin2011-1}, \cite{Callaway2011-2}, \cite{6345351}, \cite{Callaway2009}, \cite{6760990},  the optimal coordination of distributed resources in a micro-grid flexible loads \cite{5762580}, \cite{hammerstrom2007pacific}, \cite{6426122}, and real time pricing approaches to reduce peak demand \cite{conejo2010real}, \cite{6547835}, \cite{roozbehani2012volatility}. {\color{black}In contrast to DLC where consumers transfer the control of powering their appliances to a centralized decision maker, dynamic price directed demand response by individual consumers can achieve similar system benefits with lower consumer utility loss. The PNNL Olympic Peninsula project \cite{hammerstrom2007pacific1} is a successful pilot study where consumer specified price elasticities have been used to control thermostatic appliance by dynamic prices that proved superior to fixed  or peak/off-peak prices.}

{\color{black}Customer acceptance of price directed demand response, however, requires that customer preferences are modeled accurately and that dynamic prices are adapted to these preferences. In this paper we focus on the deployment of bidirectional regulation service reserves (RSR) by a pool of duty cycle cooling appliances in a smart building or neighborhood microgrid. Based up-to-date consumer preferences and the	current price signal, the microgrid operator will purchase and allocate a block or packet of energy at the beginning of each cycle. Dynamic prices encourage (or discourage) idle appliances to purchase a packet of energy and start a cycle. Two distinct time scales, one hour and 4 seconds, are involved in promising and deploying RSRs. A smart building promises to the Independent System Operator (ISO) a certain quantity of up and down RSR in the hour ahead market and is compensated at the hour ahead market’s RSR clearing price. During this hour, however, the smart building must deploy these reserves to track the RSR signal that is rebroadcast by the ISO at 4 second intervals taking values in the interval from -100\% to +100\%. Optimal deployment of the RSRs at the 4 second time scale is defined as the deployment driven by dynamic prices decided at the same 4 second time scale by the smart building operator, so as to maximize the average utility of cooling appliances over the hour. The smart building operator can not anticipate the ISO RSR signal, but it has access to a statistical model describing its stochastic dynamics during the relevant hour \cite{BilginCaramanis2012}, \cite{7346499}.
	
To obtain the optimal dynamic pricing policy, we propose and solve an infinite horizon discounted cost stochastic dynamic programming (DP) problem. The hour long horizon is considered for all intents and purposes to be infinite relative to the 4 second time scale characterizing the regulation signal and price updates. State variables consist of the RSR signal level $y(t)$, the direction $D(t)$ of RSR signal defined as the sign of $y(t)-y(t-1)$, the number of connected appliances $i(t)$, and the sufficient statistic $\hat{T}(t)$ characterizing the probability distribution of utility level. More specifically, we model the dynamics of the utility that each space conditioning appliance enjoys from the consumption of a block/packet of energy purchased when the prevailing temperature is at a specific level $T\in [T_{\min}, T_{\max}]$ within the appliance specific temperature comfort zone. The utility at time $t$ is modeled as an affine function of the temperature $T$ prevailing at time $t$. The utility of purchasing a packet of energy by a cooling appliance will be low when its temperature is close to $T_{\min}$, and high when it is close to $T_{\max}$. To model diversity among appliances, coefficients are appliance specific independent random variables and $Τ$ is normalized by $|T_{\max} - T_{\min}|$. The variable $T$ at a specific appliance evolves dynamically as a function of that appliance's past trajectory of energy block purchases and heat losses, and hence $T$ and $U(T)$ are time varying. The optimal pricing policy, however, does not require individual appliance utility levels at time $t$. The sufficient information is captured by the probability distribution of utility levels across the ensemble of all idle appliances at time $t$. This probability distribution is therefore the sufficient statistic that replaces appliance specific utilities in the stochastic DP state vector. 

The principal contribution in what follows is a modeling extension in which the system state is a time varying probability density function that we argue provides important information about the dynamics of aggregated effects of individual consumer utility functions.  This represents a significant improvement over the more traditional models (e.g. \cite{BilginCaramanis2012}, \cite{LiCaramanis2011}, \cite{7346499}) that assume a virtually infinite pool of homogeneous users and appliances whose preferences are static and independent of the recent history of price control signals.  Our approach provides a more appropriate level of granularity that will enhance the value of RSRs that come from finite pools of duty-cycle appliances.  The concepts to be described may also have useful applications in related areas such as dynamic pricing policies in the context of Internet service and mobile telephony bandwidth management \cite{Kelly1997}, \cite{Kelly1998rate}, \cite{PaschalidisTsitsiklis2000}, \cite{MutluAlanyaliStarobinskl2000}.  The bottom line message of the paper is that in communication between consumers and service providers, it is important to stop relying on the uniform and time invariant assumption and thereby avoid ratepayer revolt \cite{RatepayerRovolt}.

Focusing on a smart building with a finite number of duty cycle cooling appliances, we use a dynamic probability distribution to represent cooling zone specific occupant preferences to transition their appliance from an idle to an active state. This is achieved by employing a dynamically evolving sufficient statistic that captures the ensemble of idle cooling zone appliance preferences, namely the recent-price-trajectory-dependent probability distribution of a sampled cooling zone occupant's reservation price.} Moreover, we improve the tractability of the DP formulation by (i) deriving an analytic characterization of the optimal policy and the differential cost function, and (ii) proving useful monotonicity and convexity properties. These properties motivate the use of appropriate basis functions to construct a parametrized analytic approximation of the value function used to design an approximate DP (ADP) algorithm \cite{Bertsekas2007} that estimates optimal value function approximation parameters and near optimal policies. {\color{black}To support the relative advantage of our approach, we compare the real time RSR tracking performance of  our dynamic utility based optimal dynamic pricing policy  to that of a uniform static utility based dynamic pricing policy. The tracking of our policy will be shown to be clearly superior.}

The rest of the paper proceeds as follows. In Sec.\ref{problem formulation} we establish a DP problem by describing the state dynamics, the sufficient statistic of cooling zone preferences, the period costs, and the Bellman Equation. Time is quantized into 4-second intervals, and the electricity used by the appliances is also quantized using our previously announced concept of packetized energy. Sec.\ref{section 3} compares the static uniform distribution representation of cooling zone preferences adopted in past work, with the proposed dynamic and non-uniform preference distribution formulation. More specifically, it uses the Bellman equation to prove analytical expressions relating the optimal policy to partial differences in the value function. Sec.\ref{section:Optimal dynamic policies} proves monotonicity properties of the differential cost function and the optimal policy. It also discusses asymptotic behaviour of optimal policy sensitivities as the number of appliances becomes large. Sec.\ref{approximate dp algorithm} utilizes the proven properties of the value function and the optimal control policy to propose and implement (i) an assisted value iteration algorithm, and (ii) a parametrized approximation of the value function. It also develops an ADP approach for estimating the value function approximation parameters. {\color{black}Numerical results demonstrate the superior performance of dynamic utility versus static utility modeling pricing policies.} Sec.\ref{conclusion} concludes the paper and proposes future directions.  

\section{Problem Formulation}
\label{problem formulation}
{\color{black}
We consider an advanced energy management building with $N$ cooling appliances. The smart building operator (SBO) contracts to regulate in real time its electricity consumption within an upper and a lower limit $\{ \bar{n}-R, \bar{n}+R \}$ agreed upon in the hour-ahead market, where $\bar{n}$ is the constant energy consumption rate that the SBO purchased in the hour ahead market, and $R$ is the maximum up or down RSRs that the SBO agreed to provide in the hour-ahead market. $\bar{n}$ equals the average number of active appliances, which is related to the average number of duty cycles per hour required to maintain the cooling zone temperature within its comfort zone for the prevailing outside temperature and building heat transfer properties. $R$ is chosen as a fraction of $N$ that can be determined based on the appliance response property, such as cooling rate. In real time, the ISO would broadcast the RSR signal $y(t)$ to the SBO, which is the output of a pre-specified proportional integral filter of observed Area Control Error (ACE) and System Frequency excursions. The RSR signal $y(t)$ has an average value of 0 during each hour and its dynamics can be modeled as a Markov process \cite{PJMManual2012}. Given these three values, the SBO assumes the responsibility to modulate its energy consumption to track $\bar{n}+y(t)R$ with $y(t)\in[-1,+1]$ specified by the ISO in almost real time (usually every 2 or 4 seconds). To this end the SBO broadcasts a real time price signal $\pi(t)$ to all cooling appliances to modulate the number of connected appliances $i(t)$ and hence control the aggregated consumption. Appliances will detect price at rate $\lambda$ to resume cooling by comparing its utility $U(T)$ to $\pi(t)$.\footnote{In this paper we assume consumers to have the same sensitivity preference across zones, namely parameter $b$ in the utility function is fixed such that the mapping between temperature threshold and the price signal is bijective. However, $T_{\min}$ and $T_{\max}$ are not the same for all consumers. $|T_{\min}-T_{\max}|$ is normalized by the sensitivity parameter $b$. Alternative implementations in the literature rely on similar approaches, but, in our opinion, less likely to be accepted by consumers. For example, the PNNL Olympic Peninsula project [19], has assumed the presence of aggregators, who, typically provide participants with multiple comfort level choices, and participants choose the comfort level based on their own preferences. Aggregators group participants into multiple subgroups based on their choices. Different groups of participants are characterized by different price elasticities and provide different type of reserves. For example, a participant group with the highest elasticity provides regulation service reserves (RSR) with a response time of 4 seconds. Participant groups with medium price elasticity provide spinning reserves with response times of 15 to 30 minutes. Participant groups with low price elasticity provide capacity reserves that have even slower response times.} Appliances utility values depend on their corresponding cooling zone temperature $T$. In this paper $U(T)$ is a monotonically increasing function of $T$ over $T\in[T_{\min},T_{\max}]$. Once $U(T)>=\pi(t)$, the appliance will connect and consume an packet of energy having departure rate $\mu$. ($1/\mu$ equals the average time it takes a cooling appliance to complete the consumption of an energy packet.) See \cite{BowenBaillieul2012}, \cite{gelenbe2012energy} where the associated electric power consumption is defined as a packet of electric energy needed to provide the work required by a single cooling cycle. 

The RSR signal tracking error, $e(t)=i(t)-\bar{n}-y(t)R$, and occupant utility realizations constitute period costs. The objective is to find a state feedback optimal policy that minimizes the associated infinite horizon discounted cost. Individual cooling zone preferences are modeled by a dynamically evolving probability distribution of idle appliance zone temperatures $p_{t}(T)$. Assuming a known relationship connecting cooling zone preferences to temperatures, we derive the probability distribution of preferences $p_{t}(T)$ (or derived utility levels) for cooling appliance activation as well as the sufficient statistics $\hat{T}(t)$ that characterizes $p_{t}(T)$. We next describe system dynamics, and formulate the period cost function of the relevant stochastic dynamic problem. 
}
\subsection{State Dynamics}
The state variables contain $i(t), y(t), D(t)$ and $\hat{T}(t)$. Queues $i(t)$ and $N-i(t)$ constitute a closed queuing network where the service rate of one queue determines the arrival rate into the other. Queue $N-i(t)$ behaves like an infinite server queue with each server exhibiting a stochastic Markov modulated service rate that depends on the control $u(t)$ and the probability distribution $p_t(T)$. Queue $i(t)$ behaves also as an infinite server queue with each server exhibiting a constant service rate $\mu$. The dynamics of $y(t)$ and the dependent state variable $D(t)=\mathrm{sgn}[y(t)-y(t-4sec)]$ are characterized by transitions taking place in short but constant time intervals, $\tau_{y}$\footnote{This varies across ISOs. In PJM it is either 2 or 4 seconds depending on the type of regulation service offered.}, resulting in $y(t)$ staying constant, increasing or decreasing by a typical amount of $\Delta{y}=\tau_{y}/300 sec$ \cite{PJMManual2012}. These transitions are outputs of a proportional integral filter operated by the ISO whose inputs are system frequency deviations from 60 hz and Area Control Error (ACE). Since the frequency deviation and ACE signal can be approximated by a white noise process resulting from imbalance between stochastic demands and supply, $y(t)$ is then an unanticipated random variable which is described by memoryless transitions that depends only on the current value. Therefore we can approximate $y(t)$ by a continuous time jump Markovian process that allows us to uniformize the DP problem formulation. To uniformize the DP problem we introduce a control update period of $\Delta_{t}<<\tau_{y}$ which assures that during the period $\Delta_{t}$, the probability that more than one event can take place is negligible. We further set the time unit so that $\Delta_{t}=1$, and scale transition rate parameters accordingly and derive the following state dynamics.

\subsubsection{Dynamics of $y(t)$}
{\color{black}The transition probabilities of the discrete time Markov process $y(t)$ depend on $y(t)$ as well as $D(t)$ \cite{BilginCaramanis2012}, \cite{7346499}, which is $y(t)$'s changing direction, either positive or negative, for every 4 seconds. $D(t)=\textrm{sgn}[y(t)-y(t - 4 sec)]$.} Statistical analysis on historical PJM data on $y(t)$ trajectories indicate a weak dependence on $y(t)$ yielding the reasonable approximation:
\begin{displaymath}
\left\{ \begin{array}{l}
\mathrm{Prob}(y(t+\tau_{y})=y(t)+\Delta {y}|D(t)= 1)=0.8\\
\mathrm{Prob}(y(t+\tau_{y})=y(t)-\Delta {y}|D(t)= 1)=0.2\\
\mathrm{Prob}(y(t+\tau_{y})=y(t)-\Delta {y}|D(t)= -1)=0.8\\
\mathrm{Prob}(y(t+\tau_{y})=y(t)+\Delta {y}|D(t)= -1)=0.2
\end{array}\right. .
\end{displaymath}
Denoting by $\gamma_{1}^{u}$ ($\gamma_{1}^{d}$) the rate at which $y(t)$ will jump up by $\Delta y$ during a control update period when $D(t)=1$ ($D(t)=-1$), and by $\gamma_{2}^{u}$ ($\gamma_{2}^{d}$) the corresponding rate that $y(t)$ will jump down when $D(t)=1$ ($D(t)=-1$), and selecting these rates to correspond to the exponential rates that are consistent with the geometric probability distribution described above we have:
\begin{displaymath}
\left\{ \begin{array}{l}
\gamma_{1}^{u}~=~0.8\Delta_{t}/\tau_{y}, \gamma_{2}^{u}~=~0.2\Delta_{t}/\tau_{y}\\
\gamma_{2}^{d}~=~0.8\Delta_{t}/\tau_{y}, \gamma_{1}^{d}~=~0.2\Delta_{t}/\tau_{y}
\end{array}\right.
\end{displaymath}
Setting the control update period as the operative time unit, i.e., $\Delta_{t}=1$, we have the following uniformized dynamics of $y(t)$:
\begin{displaymath}
\left\{ \begin{array}{l}
\mathrm{Prob}(y(t+1)=y(t)+\Delta {y}|D(t)= 1)= \gamma_{1}^{u}\\
\mathrm{Prob}(y(t+1)=y(t)-\Delta {y}|D(t)= 1)= \gamma_{2}^{u}\\
\mathrm{Prob}(y(t+1)=y(t)-\Delta {y}|D(t)= -1)= \gamma_{2}^{d}\\
\mathrm{Prob}(y(t+1)=y(t)+\Delta {y}|D(t)= -1)= \gamma_{1}^{d}
\end{array}\right. .
\end{displaymath}

\subsubsection{Dynamics of $i(t)$}
The dynamics of $i(t)$ is governed by the following arrival and the departure rates. 

The arrival rate $a(t)$ depends on the policy $u(t)$. Denote by $p_{u(t)}$ the proportion of idle appliances with cooling zone temperature $T\geq u(t)$. As described in the notation subsection, idle appliances observe the price broadcast by the SBO at a rate $\lambda$, and decide to connect and resume cooling when the price is smaller than their utility for cooling at time $t$. Therefore the arrival rate into $i(t)$ is
\begin{equation}
a(t)=[N-i(t)]\lambda p_{u(t)}=(N-i(t))\lambda\int\limits_{u(t)}^{T_{\max}}p_{t}(T)dt,
\end{equation}
namely $a(t)$ equals the product of the number of idle appliances that observe the broadcast price times the probability that $T\geq u(t)$.

The departure rate $d(t)$ is independent of $u(t)$, It equals the product of active appliances times the departure rate of a cooling cycle. Modeling the consumption of an energy packet as an exponential random variable with rate $\mu$ such that $1/\mu$ equals the average cooling cycle duration, the departure rate is
\begin{equation}
d(t)=i(t)\mu.
\end{equation}
The stochastic dynamics of $i(t)$ in the homogenized model is thus given by $i(t+1)=i(t)+\tilde{i}$ where the random variable $\tilde{i}$ satisfies the following probability relations
\begin{displaymath}
\left\{ \begin{array}{l}
p(\tilde{i}=1)=a(t)\\
p(\tilde{i}=-1)=d(t)\\
p(\tilde{i}=0)=1-a(t)-d(t)- \gamma,
\end{array}\right.
\end{displaymath}
where $\gamma=\gamma_{1}^{u}+\gamma_{2}^{u}=\gamma_{1}^{d}+\gamma_{2}^{d}$ is the total probability that the ISO RSR signal will change.

\begin{figure}[hbt]
\centering
\includegraphics[width=8cm]{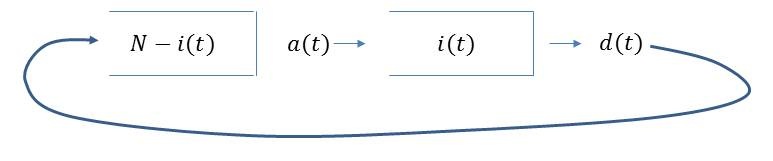}
\caption{Dynamics of the closed queuing network with infinite number of servers. The service rate at the active appliance queue is constant, while it is Markov modulated at the idle appliance queue.}
\label{constant inventory}
\end{figure}

Fig.~\ref{constant inventory} represents the stochastic dynamics of the number of active and idle appliances as a two queue closed queuing network with queue levels summing to $N$.

\subsubsection{Dynamics of $D(t)$}
Recalling that $D(t)={\rm sgn}[y(t)-y(t-4sec)]$, it is clear that the dynamics of $D(t)$ are fully determined by the dynamics of $y(t)$. We next show that the dynamics of $\hat{T}(t)$ are also determined by the dynamics of $y(t)$.

\subsubsection{Dynamics of $\hat{T}(t)$}
\begin{figure}[hbt]
	\centering
	\includegraphics[width=9cm]{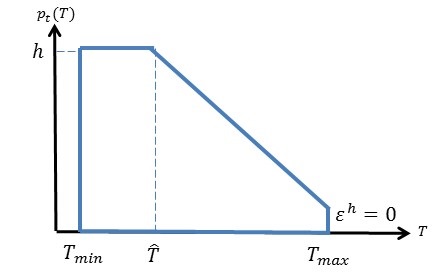}
	
	\caption{Trapezoid probability distribution function $p(T)$ with $T\in[T_{\min},T_{\max}]$ parametrized by a single parameter $\hat{T}$. {\color{black} Simulation shows that $\epsilon^{h}$ is of small value and assumed to be zero throughout paper.} The height of the trapezoid is $h=2/(T_{\max}+\hat{T}-2T_{\min})$.}
	\label{pdf_orig}
\end{figure}

Based on related work in HVAC system modelling \cite{Chassin2011-1} and \cite{Callaway2012}, we use standard energy transfer relations to simulate the dynamics of the frequency histogram of idle appliance cooling zone temperatures, which appear to conform to a three parameter functional representation $p_{t}(T)=f(\hat{T}(t),T_{\min},T_{\max})$. We simulate two RSR signals from the ISO for the observation of the time varying probability distribution of consumers' temperature. The first is a standard ISO RSR signal trajectory that aspiring RSR market participants must respond to in order to demonstrate that they have the ability to track. This is referred to in the PJM manual as the standard T-50 qualifying test \cite{PJMManual2012}. The second is a real time signal downloaded from PJM \cite{PJMSignal2012}. We record the temperature levels prevailing across the $N$ cooling zones when a control trajectory is applied that results in near-perfect tracking of the ISO RSR requests implied by the aforementioned two signals. {\color{black}Assuming standard heat transfer relationships, extensive simulation reported below indicates that the time evolution of the probability distribution of cooling zone temperatures conforms to a dynamically changing trapezoid characterized fully by $\hat{T}(t)$, see Fig.~\ref{pdf_orig}. $p_t(T)$ is the probability density function (pdf) of $T$ at time $t$, $T\in[T_{\min},T_{\max}]$. $T_{\min}$ and $T_{\max}$ are the threshold values of room temperature that determine the appliance cooling zone occupant's comfort zone. $\epsilon^{h}$ is of small value and approximated to be zero.} Trapezoid has base $(T_{\min},0)$ to $(T_{\max},0)$ and top side $(T_{\min},h)$ to $(\hat{T},h)$ where $h=2/(T_{\max}+\hat{T}-2T_{\min})$. Note that $\hat{T}$ and hence $h$ are time varying quantities. We will using $p(T)$ to denote the pdf by dropping the time index in derivation and proof if time is not considered.

\begin{figure}[hbt]
\centering
\includegraphics[width=9cm]{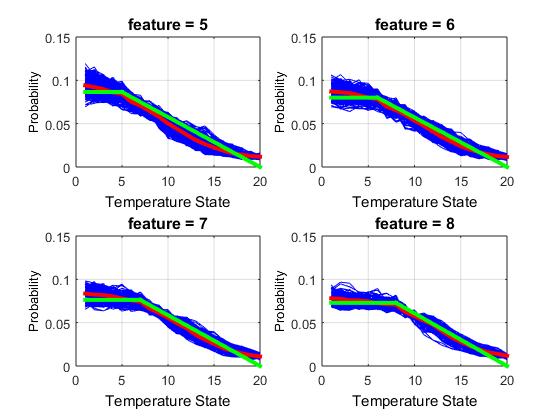}

\caption{Trapezoid probability distribution function $p(T)$ with $T\in[T_{\min},T_{\max}]$ parametrized by a single parameter $\hat{T}$. Height of the trapezoid is $h=2/(T_{\max}+\hat{T}-2T_{\min})$.}
\label{statistics_of_the_trapezoid}
\end{figure}

Fig. \ref{statistics_of_the_trapezoid} shows the accuracy of using a trapezoid probability distribution to represent idle appliance cooling zone temperatures obtained from a Monte Carlo Simulation of the price controlled system. We discretize the temperature into 20 states and simulate a total number of 16000 appliances to observe a smooth probability distribution. The duty cycle on and off time are both 10 minutes. The price detection rate from idle appliances is 1 minute. PJM's RSR signal is broadcast every 4 seconds. In Fig. \ref{statistics_of_the_trapezoid}, the numerous black curves are the actual probability distribution recorded at different time stamps for trapezoids characterized with $\hat{T}(t)=5,6,7,8$. The red curve is the mean value of the set of black curves taken at each temperature state. The red curve is then approximated by the trapezoid green curve proposed in Fig.~\ref{pdf_orig}. $\hat{T}(t)$ and the trapezoid approximation are then the mean statistics of the actual frequency distribution of the temperature distribution. Note that the trapezoids are completely specified by two static quantities, $T_{\min}$ and $T_{\max}$, forming the base of the trapezoid, and the time varying quantity $\hat{T}(t)$ that determines the height of the top horizontal side. {\color{black}It should be noted that the trapezoid approximation is based on the assumption of (1) small value of $\epsilon^{h}=0$ and (2) equal probability $p_{t}(T)$ for temperature $T\in[T_{\min},\hat{T}]$. In this paper, we model the trapezoid distribution based on the single sufficient statistics $\hat{T}$, future work will consider modeling the preference distribution with additional statistics to fully capture the distribution shape.}

\begin{figure}[hbt]
\centering
\includegraphics[width=9cm]{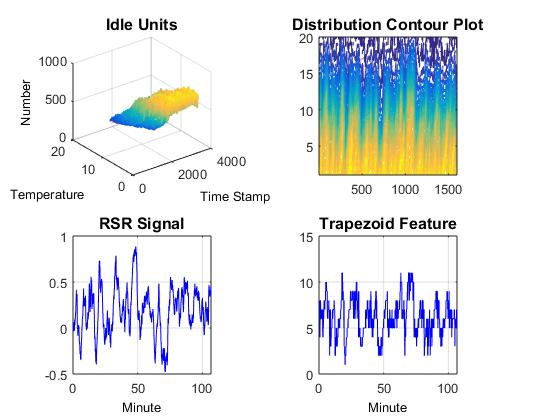}

\caption{Monte Carlo simulation is used to determine the time evolution of temperature histograms in idle appliance cooling areas. We observe that these histograms conform to a time varying trapezoid shape with $\hat{T}(t)$ a linear function of the ISO RSR signal $y(t)$.}
\label{evolution_of_the_preferences}
\end{figure}

\begin{figure}[hbt]
\centering
\includegraphics[width=9cm]{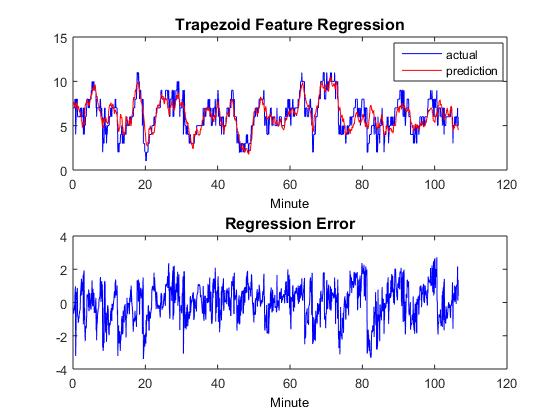}

\caption{Original $\hat{T}$ and the regression based prediction of $\hat{T}$ are shown in the upper figure. The linear regression explains 81.06\% of the uncertainty of $\hat{T}$. Errors are observed to be Gaussian noise satisfying the assumption of regression.}
\label{fitted_curve_oft_hat}
\end{figure}

The upper left plot in Fig. \ref{evolution_of_the_preferences} shows the time evolution of the trapezoids representing the idle appliance cooling zone temperature histograms throughout time when we are simulating the RSR tracking for the real signal. The upper right plot is the contour plot of the number of appliances where we can see clearly a time varying trapezoid distribution shaping the preferences. We filter the contour plot to get $\hat{T}$ in lower right figure. Based on the time series recorded for $y(t)$ (lower left figure) and $\hat{T}(t)$, we find a strong anti-correlation between the two vectors in both simulation for the T-50 standard signal and the real RSR signal that are given by -0.9833 and -0.8106, respectively. We propose the regression of $\hat{T}(t)$ on $y(t)$ with the following linear function
\begin{equation}\label{dynamics of hat t}
\hat{T}(t)=\alpha_{0}+\alpha_{1}y(t)+\omega,
\end{equation}
where $\alpha_{0}$ corresponds to the value of $\hat{T}(t)$ when the building's energy consumption level is $\bar{n}$, $\alpha_{1}<0$, and $\omega$ is a zero mean symmetrically distributed error. These results not only explain most of the variability but are also sensible and conform with our expectations. Indeed, large values of $y(t)$ indicate a history of repeated broadcasts of low prices to achieve high consumption levels requested by the ISO. Small values of $\hat{T}(t)$ approaching $T_{\min}$ are observed for high $y(t)$ levels, while for low levels of $y(t)$, $\hat{T}(t)$ is large. These findings support our a priori expectation that $y(t)$ is a reasonable sufficient statistic of past state and control trajectories in the information vector available at time $t$. This a priori expectation is based on the fact that $y(t)$ levels are in fact integrators of recent price control trajectories. The verification of expectations by actual observations justifies the adoption of a tractable dynamic utility model conforming to the dynamics $\hat{T}(t)=\alpha_{0}+\alpha_{1}y(t)+\omega$ where $\omega$ is a zero mean symmetric random variable. Since the SBO is able to observe the actual cooling zone temperatures through its access to Building Automation Control (BAC), the dynamics above are adequate for optimal control estimation. We finally note that the mapping of temperature to consumption utility allows the dynamic and past-control-dependent distribution of cooling area zone temperatures to provide a dynamic and past-control-dependent distribution of cooling area consumption utility levels. In the end, Fig.~\ref{fitted_curve_oft_hat} shows the actual $\hat{T}$ (black curve), the predicted $\hat{T}$ (red curve) based on linear regression, and the error between the two values. It can be seen that the error is zeros mean and symmetrically distributed that is consistent with our assumption in proposing (\ref{dynamics of hat t}).

\subsection{Period Cost}
The period cost consists of two parts: a penalty for deficient ISO RSR signal tracking and the utility realized by appliance users. The penalty for deficient tracking rate at time $t$ is defined as:
\begin{equation}\label{kappa penalty}
g(i(t),y(t))=K[\frac{i(t)-\bar{n}-y(t)R}{R}]^{2},
\end{equation}
where $K$ is the penalty per unit of deficient tracking. Defining $\kappa=K/R^{2}$ we can write the penalty for deficient tracking as,
\begin{equation}\label{period penalty cost}
g(i(t),y(t))=\kappa[i(t)-\bar{n}-y(t)R]^{2}.
\end{equation}
{\color{black}The expected utility realized from a group of idle appliances can be calculated by three components: (1) the number of idle appliances, $N-i(t)$, that can potentially connect, (2) the connection probability of each appliance, $p_{u(t)}$, that depends on the dynamic frequency distribution $p_{t}(T)$, and (3) the utility rate per connection, which is the expected monetary value per connection calculated by consumers' temperature based utility function $U(T)$ and the probability distribution of room temperature $T$. It should be noted that while the utility for a given consumer is determined by his room temperature $T$, for a group of consumers the aggregated utility is based on the frequency distribution of room temperatures of the group, and therefore depends on $p_{t}(T)$ instead of $T$.

For a given price signal $\pi(t)$, there is a threshold temperature value $u(t)$ obtained by solving $\pi(t)= U(u(t))$. Idle appliances having temperature at least $u(t)$ will connect. The consequent expected utility value realized by each connecting cooling appliance is
\begin{equation}\label{expected utility for one person}
U_{u(t)}=\frac{\int\limits_{u(t)}^{T_{\max}}U(T)p_{t}(T)dT}{\int\limits_{u(t)}^{T_{\max}}p_{t}(T)dT}.
\end{equation}
Denoting $a(t)$ as the expected number of connections from the idle appliance group, the period expected utility value is}
\begin{equation}
\label{utility}
\begin{array}{lll}
a(t)U_{u(t)} & = & [N-i(t)]\lambda p_{u(t)}U_{u(t)},\\
& = & [N-i(t)]\lambda\int\limits_{u(t)}^{T_{\max}}p_{t}(T)dt \frac{\int\limits_{u(t)}^{T_{\max}}U(T)p_{t}(T)dT}{\int\limits_{u(t)}^{T_{\max}}p_{t}(T)dT},\\
& = &[N-i(t)]\lambda \int\limits_{u(t)}^{T_{\max}}U(T)p_{t}(T)dT.
\end{array}
\end{equation}
Equations (\ref{period penalty cost}) and (\ref{utility}) imply that the total cost rate is
\begin{equation}
\begin{array}{lll}
c(i(t),y(t),u(t))&=&\kappa[i(t)-\bar{n}-y(t)R]^{2}-\\
&&[N-i(t)]\lambda \int\limits_{u(t)}^{T_{\max}}U(T)p_{t}(T)dT.
\end{array}
\end{equation}

\subsection{Bellman Equation}
The state variables can be grouped according to their dependence on $u(t)$: $i(t)$ depends explicitly on $u(t)$. $\hat{T}(t)$ is also dependent on the past trajectory of controls, but, to the extent that this trajectory is consistent with a reasonable tracking the ISO RSR signal, it can be considered as a function of $y(t)$, which, as discussed earlier, is the sufficient statistic of this trajectory. We can thus consider all state variables, other than $i(t)$, to have dynamics that do not depend on $u(t)$. For notation simplicity we let $\bar{i}^{u}(t)=\{y(t),D(t)=1,\hat{T}(t)\}$ ($\bar{i}^{d}(t)=\{y(t),D(t)=-1,\hat{T}(t)\}$) to be the state variables that make up the complement of $i(t)$ when the RSR signal is going up (down), so that $\{i(t),\bar{i}^{u(d)}(t)\}$ is the representation of the full state vector. Given the cost function and dynamics described above, we can formulate an infinite horizon discounted cost problem with the following Bellman equation for states including $D(t) = -1$.
\begin{eqnarray}\label{Bellman equation}
J(i,\bar{i}^{d}) &=& \min_{u\in[T_{\min},T_{\max}]} \big\{g(i,\bar{i}^{d})-a(t)U_{u}\nonumber\\ 
&& +\alpha[a(t)J(i+1,\bar{i}^{d}) + d(t)J(i-1,\bar{i}^{d})\nonumber\\ 
&& +\gamma_{1}^{d}J(i,\bar{i}^{u} + \Delta {y}) + \gamma_{2}^{d}J(i,\bar{i}^{d}-\Delta {y}) \nonumber\\
&& +(1-a(t)-d(t)-\gamma_{1}^{d}-\gamma_{2}^{d})J(i,\bar{i}^{d})]\}.
\end{eqnarray}
$J(i,\bar{i}^{d})$ is the value function satisfying the Bellman equation, with $\alpha$ denoting the discount factor. For notational simplicity we denote by $\bar{i}^{u}+\Delta {y}$ the new state realized when the regulation signal increases from $y(t)$ to $y(t+1) = y(t)+ \Delta {y}$ rendering $D(t+1)=1$, while the rest of the state variables remain unchanged. Similarly we denote by $\bar{i}^{d}-\Delta {y}$ the new state when the regulation signal decreases from $y(t)$ to $y(t+1) = y(t)- \Delta {y}$ rendering $D(t+1)=-1$, while the rest state variables remain unchanged. The superscripts $u$ $(d)$ standing for upwards (downwards) RSR signals. $ J(i,\bar{i}^{u})$ can be written similarly with minor notational changes.

\section{Utility Realization and Optimal Policy}
\label{section 3}
\subsection{Uniform Utility Probability Distribution Model}
For illustration purposes, we start with a simple utility function that represents a linear relationship between cooling zone temperature and utility enjoyed by activating an idle appliance and allowing it to embark on a cooling cycle 
\begin{equation}\label{utility function}
U(T)=b(T-T_{\min}).
\end{equation}
The utility increases proportionately to the cooling zone temperature  $T$. If $p(T)$ were selected to be a static and uniform probability distribution, as is the case with work published so far, the expected period utility rate would be a conveniently concave function of $u$. Indeed, using (\ref{utility}) we can obtain that the expected period utility rate
\begin{equation}
\begin{array}{lll}
a(t)U_{u} &= & (N-i(t))\lambda\int\limits_{u}^{T_{\max}}U(T)p(T)dT,\\
& = & (N-i(t))\lambda\int\limits_{u}^{T_{\max}}b(T-T_{\min})\frac{1}{T_{\max}-T_{\min}}dT,\\
& = & (N-i(t))\lambda\frac{b(T_{\max}-u)(T_{\max}+u-2T_{\min})}{2(T_{\max}-T_{\min})},
\end{array}
\end{equation}
is a concave function of the policy $u$. 

In general, this concavity property of the expected period utility rate holds for broader class of utility functions $U(T,b,\epsilon)$ -- where $T$ is the temperature, $b$ is the utility sensitivity, and $\epsilon$ is a consumer specific value to characterize individual utility preference. We require two properties hold for the utility function: $(i)$ we need $\epsilon$ to be a random parameter independent of $T$, and $(ii)$ $E_{\epsilon}[U(T,b,\epsilon)]$ is a monotonically increasing function of $T$. We prove this property by the following reasoning. Taking expectation on $U_{u}$ in (\ref{expected utility for one person}), we have
\begin{equation}
U_{u}=E_{\epsilon}\left[\frac{\int\limits_{u(t)}^{T_{\max}}U(T,b,\epsilon)p(T)dT}{\int\limits_{u(t)}^{T_{\max}}p(T)dT}\right]= \frac{\int\limits_{u(t)}^{T_{\max}}E_{\epsilon}[U(T,b,\epsilon)]p(T)dT}{\int\limits_{u(t)}^{T_{\max}}p(T)dT}.
\end{equation}
Taking the derivative of the expected utility rate $a(t)U_{u}$ with respect to $u$
\begin{equation}
\begin{array}{lll}
\frac{d}{du}a(t)U_{u} &=& \frac{d}{du}(N-i(t))\lambda\int\limits_{u}^{T_{\max}} U(T)p(T)dT,\\
&=& -\frac{(N-i(t))\lambda}{T_{\max}-T_{\min}}E_{\epsilon}[U(u,b,\epsilon)].
\end{array}
\end{equation} 
Since $E_{\epsilon}[U(u,p,\epsilon)]$ increases with $u$, we have $\frac{d}{d u}E_{\epsilon}[U(u,p,\epsilon)]>0$, which leads to
\begin{equation}
\frac{d^{2}}{d u^{2}}a(t)U_{u}= -\frac{(N-i(t))\lambda}{T_{\max}-T_{\min}}\frac{d}{d u}E_{\epsilon}[U(u,p,\epsilon)]<0
\end{equation}
Hence the expected period utility rate is a concave function of the policy $u$.
\subsection{Generalized Utility Probability Distribution Model}
The concavity property no longer holds true under the realistic modelling of $p(T)$ by a dynamic trapezoid characterized additionally by the time varying quantity $\hat{T}$. Indeed, the realistic representation implies the following consumers' preferences distribution
\begin{displaymath}
p(T)=\left\{
\begin{array}{ll}
\frac{2}{T_{\max}+\hat{T}-2T_{\min}}, & T\leq\hat{T},\\
\frac{2(T-T_{\max})}{(\hat{T}-T_{\max})(T_{\max}+\hat{T}-2T_{\min})}, & T\geq\hat{T}.
\end{array}
\right.
\end{displaymath} 
For example, if we use a linear utility function as in (\ref{utility function}), it yields the following expected period utility rate
\begin{equation}
\label{exp. etility per connection}
\begin{array}{ll}
& a(t)U_{u}\\
= & \left\{
\begin{array}{ll}
[N-i(t)]\lambda\frac{2b(C_{1} - \frac{1}{2}u^{2}+T_{\min}u)}{T_{\max}+\hat{T}-2T_{\min}}, & u\leq\hat{T},\\[7pt]
[N-i(t)]\lambda\frac{2b( C_{2}-\frac{1}{3}u^{3}+\frac{T_{\min}+T_{\max}}{2}u^{2}-T_{\min}T_{\max}u  )}{(\hat{T}-T_{\max})(T_{\max}+\hat{T}-2T_{\min})}, & u\geq\hat{T}.
\end{array}
\right.
\end{array}
\end{equation}
where $C_{1}$ and $C_{2}$ are some constants. The introduction of a dynamic $\hat{T(t)}$ dependent $p(T)$ removes the concavity of the expected utility rate since the second derivative of the expected utility is 
\begin{equation}
\frac{d^{2}}{d u^{2}}a(t)U_{u} \propto T_{\min} + T_{\max} - 2u.
\end{equation}
And therefore the expected period utility rate is concave for $u\in[T_{\min},\max(\hat{T},\frac{T_{\min}+T_{\max}}{2})]$, and convex for $u\in[\max(\hat{T},\frac{T_{\min}+T_{\max}}{2}), T_{\max}]$. 

Under the static uniform probability distribution $p(T)$, the optimal policy can be easily obtained since we can set derivative to zero to get a local maximum which is also global. In addition, we proceed to show that a unique optimal policy exists as well under dynamic $p(T)$ circumstances. We do this by showing first that a local maximum exists, and then prove that only one local maximum exists, and hence it is the global maximum as well.

\subsection{Optimal Price Policy}
We define the differential of the value function $J(i,\bar{i}^{d})$ w.r.t. the active appliance state variable $i(t)$ as
\begin{displaymath} 
\Delta(i+1,\bar{i}^{d})=J(i+1,\bar{i}^{d})-J(i,\bar{i}^{d}).
\end{displaymath}
Using the Bellman equation, we can express the optimal policy $u(i,\bar{i}^{d})$ in terms of $\Delta(i+1,\bar{i}^{d})$
\begin{equation}
\label{arg min u}
\begin{array}{cll}
u(i,\bar{i}^{d})&=&\arg\min_{u} g(i,\bar{i}^{d})-\lambda(N-i)p_{u}U_{u}+\\
&& \alpha\big\{i\mu J(i-1,\bar{i}^{d}) + \lambda(N-i)p_{u}J(i+1,\bar{i}^{d})+\\
&& \gamma_{1}^{d}J(i,\bar{i}^{u}+\Delta y)+\gamma_{2}^{d}J(i,\bar{i}^{d}-\Delta y)\\
&& + [1-(i\mu+\lambda(N-i)p_{u}+\gamma_{1}^{d}+\gamma_{2}^{d})]J(i,\bar{i}^{d})\big\},\\
&=& \arg \max_{u} p_{u}U_{u}-\alpha p_{u}\Delta(i+1,\bar{i}^{d}),
\end{array}
\end{equation}
where the second equation is obtained by neglecting terms that are independent of $u$. Letting
\begin{equation}
f(u,\Delta(i+1,\bar{i}^{d}))=p_{u}U_{u}-\alpha p_{u}\Delta(i+1,\bar{i}^{d}),
\end{equation}
we can write that the optimal policy must satisfy 
\begin{equation}
\label{optimization of policy}
\max_{u\in[T_{\min},T_{\max}]} \hspace{5mm} f(u,\Delta(i+1,\bar{i}^{d})).
\end{equation}

\textbf{Proposition 1} In the conventional assumption where consumers' utility preference is statically uniformly distributed ($\hat{T} = T_{\max}$), $f(u,\Delta(i+1,\bar{i}^{d}))$ is a concave function of the policy $u$ in the allowable control set. The optimal policy is obtained either at the boundary of the allowable control set or at $u^{\star}$ satisfying $\frac{d}{du}f(u,\Delta(i+1,\bar{i}^{d}))|_{u=u^{\star}}=0$.

Proposition 1 is straightforward because the first term in $f(u,\Delta(i,\bar{i}^{d}))$ is quadratic and the second term is a linear function of $u$ for $\hat{T} = T_{\max}$. When $\hat{T}< T_{\max}$ with $p(T)$ no longer uniform but trapezoid, $f(u,\Delta(i+1,\bar{i}^{d}))$ no longer possesses the concavity property which under Proposition 1 guaranteed that a local maximum is the global maximum. We therefore proceed to prove existence and uniqueness as follows.

\textbf{Proposition 2} For trapezoid consumers' preference distribution $p(T)$ with $\hat{T} < T_{\max}$ and non-homogeneous preferences function $U(T,b,\epsilon)$ having monotonically increasing expected value $E_{\epsilon}[U(T,b,\epsilon)]$, the optimal policy that solves (\ref{optimization of policy}) is described by the following relations

{\small
\begin{equation}
\label{saturation control}
u(i,\bar{i}^{d})=\left\{
\begin{array}{ll}
T_{\max}, & \mathrm{if} \hspace{3mm} \alpha\Delta(i+1,\bar{i}^{d})\geq E_{\epsilon}[U(T_{\max},b,\epsilon)]\\
T_{\min}, & \mathrm{if} \hspace{3mm} \alpha\Delta(i+1,\bar{i}^{d})\leq 0\\
\textit{I}^{-1}(\alpha \Delta(i+1,\bar{i}^{d})), & \mathrm{otherwise}
\end{array}
\right. 
\end{equation}
}where $\textit{I}^{-1}(\alpha \Delta(i+1,\bar{i}^{d}))$ is the inverse function of the expected utility function satisfying $E_{\epsilon}[U(u,b,\epsilon)]=\alpha \Delta(i+1,\bar{i}^{d})$.

\textit{Proof.} Let $f(u,\Delta(i+1,\bar{i}^{d})) = p_{u}(U_{u}-\alpha\Delta(i+1,\bar{i}^{d}))$, for the three conditions in (\ref{saturation control}) we claim the following:

\textit{1)} When $\alpha\Delta(i+1,\bar{i}^{d})\geq E_{\epsilon}[U(T_{\max},b,\epsilon)]$, namely $\alpha\Delta(i+1,\bar{i}^{d})$ is no less than the maximum possible utility per connection, we always have $U_{u}-\alpha\Delta(i+1,\bar{i}^{d})\leq 0$. Since $U_{u}-\alpha\Delta(i+1,\bar{i}^{d})$ is a monotonically increasing function and $p_{u}\geq 0$ is a monotonically decreasing function of $u$, $f(u,\Delta(i+1,\bar{i}^{d}))$ reaches its maximum value at $u=T_{\max}$. On the other hand, if $u(i,\bar{i}^{d}) = T_{\max}$ which is the optimal policy, we must have $\alpha\Delta(i+1,\bar{i}^{d})\geq E_{\epsilon}[U(T_{\max},b,\epsilon)]$. To see this necessity, assume that $\alpha\Delta(i+1,\bar{i}^{d}) < E_{\epsilon}[U(T_{\max},b,\epsilon)]$, then there exists a policy $u\neq T_{\max}$ such that $E_{\epsilon}[U(T_{\max},b,\epsilon)] > U_{u} > \alpha\Delta(i+1,\bar{i}^{d})$ and $p_{u}>0$. Hence $f(u,\Delta(i+1,\bar{i}^{d}))>0=f(T_{\max},\Delta(i+1,\bar{i}^{d}))$, which is a contradiction to the assumption that $u(i,\bar{i}^{d}) = T_{\max}$ is optimal.  

\textit{2)} When $\alpha\Delta(i+1,\bar{i}^{d})\leq 0$, both $-p_{u}\alpha\Delta(i+1,\bar{i}^{d}))$ and $p_{u}U_{u}$ are monotonically decreasing function of $u$. Therefore $f(u,\Delta(i+1,\bar{i}^{d}))$ reaches its maximum at $u=T_{\min}$. 

\textit{3)} When $\alpha\Delta(i+1,\bar{i}^{d})\in (0,E_{\epsilon}[U(T_{\max},b,\epsilon)])$, we can take the derivative of $f(u,\Delta(i+1,\bar{i}^{d}))$ as $f(u,\Delta(i+1,\bar{i}^{d}))$ is continuously differentiable on $(T_{\min},T_{\max})$. 
\begin{equation}
\label{prop 2 case 3}
\begin{array}{ll}
& \frac{\displaystyle d}{\displaystyle d u}f(u,\Delta(i+1,\bar{i}^{d})), \\
= & \frac{\displaystyle d}{\displaystyle d u} [p_{u}U_{u} - \alpha p_{u}\Delta(i+1,\bar{i}^{d})], \\
= & \frac{\displaystyle d}{\displaystyle d u} [\int\limits_{u}^{T_{\max}}E_{\epsilon}[U(T,b,\epsilon)]p(T)dT - \alpha \Delta(i+1,\bar{i}^{d}) \int\limits_{u}^{T_{\max}}p(T)dT], \\
= & -p(u)[E_{\epsilon}[U(u,b,\epsilon)]-\alpha\Delta(i+1,\bar{i}^{d})].
\end{array}
\end{equation}

Denoting $u(i,\bar{i}^{d})$ as the optimal control that minimizes $f(u,\Delta(i+1,\bar{i}^{d}))$, a necessary condition is to have $u(i,\bar{i}^{d})$ be a local maximum of $f(u,\Delta(i+1,\bar{i}^{d}))$. Therefore it satisfies the first order condition
\begin{equation}\label{new prop 2}
p(u)[E_{\epsilon}[U(u,T,\epsilon)]-\alpha\Delta(i+1,\bar{i}^{d})]\big|_{u=u(i,\bar{i}^{d})} = 0.
\end{equation}
According to the proof in \textit{1)}, $p(u)=0$ ($u(i,\bar{i}^{d}) = T_{\max}$) if and only if $\alpha\Delta(i+1,\bar{i}^{d})\geq E_{\epsilon}[U(T_{\max},b,\epsilon)]$. Therefore in this case $p(u)\neq 0$. The only solution to satisfy (\ref{new prop 2}) is 
\begin{equation}\label{eq19}
E_{\epsilon}[U(u(i,\bar{i}^{d}),b,\epsilon)]-\alpha\Delta(i+1,\bar{i}^{d})=0.
\end{equation}
Based on the definition of the inverse function, we have
\begin{equation}
u(i,\bar{i}^{d})=\textit{I}^{-1}(\alpha \Delta(i+1,\bar{i}^{d})).
\end{equation}
For the second order condition, it can be verified that $\frac{d^{2}}{d u^{2}}f(u,\Delta(i+1,\bar{i}^{d}))\big\arrowvert_{u=u(i,\bar{i}^{d})}< 0$. Therefore $u(i,\bar{i}^{d})$ is a local maximum. Moreover, given $f(u,\Delta(i+1,\bar{i}^{d}))$ is first order differentiable, $\frac{d}{du}f(u,\Delta(i+1,\bar{i}^{d}))$ is continuous and has only one critical point inside the allowable control set, then the local maximum is the global maximum for $u\in[T_{\min},T_{\max}]$, namely $u(i,\bar{i}^{d})=\textit{I}^{-1}(\alpha \Delta(i+1,\bar{i}^{d}))$. \hfill\(\Box\)

\textbf{Remark 1} The optimal policy characterization between $u(i,\bar{i}^{d})$ and $\Delta(i+1,\bar{i}^{d})$ does not rely on $p(T)$, namely it holds for broader possible realizations of consumers' real time preferences. This is because (\ref{eq19}) has only one solution which is the local and global optimal bearing the same argument in the proof. In addition, it holds for broader class of utility function (linear, quadratic, etc) and non-homogeneous utility incorporating consumer specific preference $\epsilon$.

\textbf{Remark 2} The optimal policy is determined by balancing (1) the utility rewards from connected consumers, and (2) the differential of the optimal cost viewed as an estimate of the value function difference across two adjacent states. Consumers utility sensitivity $b$ plays the following role: When $b$ increases, then the optimal policy will decrease for the same value of $\Delta(i+1,\bar{i}^{d})$. In the extreme case when $b\rightarrow\infty$, we have $u = T_{\min}$ namely the lowest price is broadcast to guarantee the largest possible utility reward; when $b\rightarrow 0$, the optimal controller is bang-bang depending on the sign of $\Delta(i+1,\bar{i}^{d})$ indicating that consumers become extremely elastic.

\begin{figure}[hbt]
\centering

\includegraphics[width=9cm]{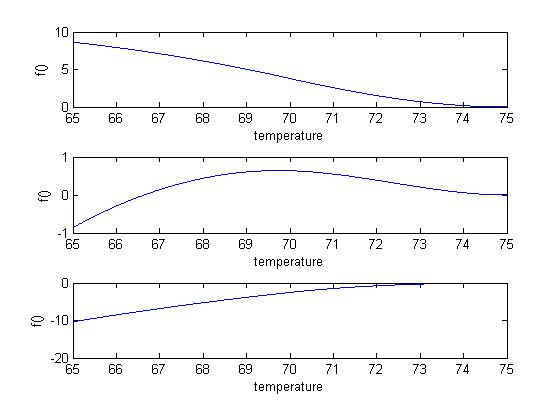}

\caption{Geometric features of $f(u,\Delta(i,\bar{i}^{d}))$. (1) $f(u,\Delta(i,\bar{i}^{d}))$ is monotonically decreasing when $\alpha\Delta(i+1,\bar{i}^{d})\leq 0$. Optimal policy is $u=T_{\min}$. (2) $f(u,\Delta(i,\bar{i}^{d}))$ has unique global maximum when $\alpha\Delta(i+1,\bar{i}^{d})\in (0,E_{\epsilon}[U(T_{\max},b,\epsilon)])$. The function may not be concave for $\hat{T}\neq T_{\max}$. (3) $f(u,\Delta(i,\bar{i}^{d}))$ is monotonically increasing when $\alpha\Delta(i+1,\bar{i}^{d})\geq E_{\epsilon}[U(T_{\max},b,\epsilon)]$. Optimal policy is $u=T_{\max}$.}
\label{three possible function to be optimized}
\end{figure}

\textbf{Remark 3} The three cases in Proposition 2 correspond to different geometry of $f(u,\Delta(i+1,\bar{i}^{d}))$; see Fig.~\ref{three possible function to be optimized}. With different choice of utility function and $\Delta(i+1,\bar{i}^{d})$, $f(u,\Delta(i+1,\bar{i}^{d}))$ can be a monotonically increasing function of $u$ that leads to the optimal control $u(i,\bar{i}^{d})=T_{\max}$, or it can be a monotonically decreasing function to render $u(i,\bar{i}^{d})=T_{\min}$, or can be a non-concave and non-monotonic function whose local maximum is the global maximum on $(T_{\min},T_{\max})$.

\section{Properties of the Optimal Policy}\label{section:Optimal dynamic policies}
Proposition 2 expresses the optimal policy $u(i,\bar{i}^{d})$ as a function of $\Delta(i,\bar{i}^{d})$. To study the properties of $u(i,\bar{i}^{d})$, we focus on the structure of $\Delta(i,\bar{i}^{d})$. In this section we derive key properties of $\Delta(i,\bar{i}^{d})$ in terms of the changes in state space variables that lead to desirable structures for $u(i,\bar{i}^{d})$. There are three state variables that affect $u(i,\bar{i}^{d})$: the aggregate consumption over all active appliances $i(t)$, the ISO RSR signal $y(t)$, and the tracking error $e(t)=i(t)-\bar{n}-y(t)R$. When two of the three variables are given, the third variable can be expressed accordingly by $i(t) - y(t)R  - e(t) = \bar{n}$ since $\bar{n}$ is fixed. To study the structure of $\Delta(i,\bar{i}^{d})$ as a function of $i(t)$, $y(t)$, and $e(t)$, we fix one variable each time and allow the other two to vary. Before proceeding, we prove the following:

\textbf{Lemma 1} Denote 
\begin{equation}
\phi(\Delta(i+1,\bar{i}^{d}))=\max_{u\in[T_{\min},T_{\max}]} p_{u} U_{u}-\alpha p_{u}\Delta(i+1,\bar{i}^{d}),
\end{equation}
then $\phi(\Delta(i+1,\bar{i}^{d}))$ is a monotonically non-increasing function.

\textit{Proof.} For saturated optimal control $u(i,\bar{i}^{d})=T_{\min}$ or $u(i,\bar{i}^{d})=T_{\max}$, $p_{u(i,\bar{i}^{d})}$ and $U_{u(i,\bar{i}^{d})}$ are constant and the statements stand. When the optimal control is not saturated, namely $u(i,\bar{i}^{d})=\textit{I}^{-1}(\alpha \Delta(i+1,\bar{i}^{d}))$ as in the last scenario in Proposition 2, we have
\begin{displaymath}
\begin{array}{ll}
& \frac{d\phi(\Delta(i+1,\bar{i}^{d})) }{d \Delta(i+1,\bar{i}^{d})}\\
=& \left[\frac{d p_{u}U_{u}}{d u}|_{u=u(i,\bar{i}^{d})} - \alpha\Delta(i+1,\bar{i}^{d})\frac{d p_{u}}{d u }|_{u=u(i,\bar{i}^{d})}\right]\frac{d u}{d \Delta(i+1,\bar{i}^{d})}\\
& -\alpha p_{u(i,\bar{i}^{d})}\\
=& -\left[E[U(u(i,\bar{i}^{d}),b,\epsilon)]-\alpha\Delta(i+1,\bar{i}^{d})\right] p(u(i,\bar{i}^{d}))\frac{d u}{d \Delta(i+1,\bar{i}^{d})}\\
& - \alpha p_{u}\\
=&-\alpha p_{u} \leq 0.
\end{array}
\end{displaymath}
Therefore $\phi(\Delta(i+1,\bar{i}^{d}))$ is a monotonically non-increasing function of $\Delta(i+1,\bar{i}^{d})$.\hfill\(\Box\)

In addition to the monotonicity properties of $\phi(\Delta(i+1,\bar{i}^{d}))$, we derive upper and lower bounds on the change in $\phi(\Delta(i+1,\bar{i}^{d}))$ with respect to a change in $\Delta(i+1,\bar{i}^{d})$ shown in the following Lemma:

\textbf{Lemma 2} $\phi(\Delta(i+1,\bar{i}^{d}))-\phi(\Delta(i,\bar{i}^{d}))$ has the following upper and lower bound:

(1) $\phi(\Delta(i+1,\bar{i}^{d}))-\phi(\Delta(i,\bar{i}^{d}))\leq-\alpha p_{u(i,\bar{i}^{d})}(\Delta(i+1,\bar{i}^{d})-\Delta(i,\bar{i}^{d}))$.

(2) $\phi(\Delta(i+1,\bar{i}^{d}))-\phi(\Delta(i,\bar{i}^{d}))\geq-\alpha p_{u(i-1,\bar{i}^{d})}(\Delta(i+1,\bar{i}^{d})-\Delta(i,\bar{i}^{d}))$.

\textit{Proof.} (1). The proof is straightforward
\begin{equation}
\begin{array}{ll}
& \phi(\Delta(i+1,\bar{i}^{d}))-\phi(\Delta(i,\bar{i}^{d}))\\
= & [p_{u}U_{u}-\alpha p_{u}\Delta(i+1,\bar{i}^{d})] \big\arrowvert_{u=u(i,\bar{i}^{d})}\\
& -[p_{u}U_{u}-\alpha p_{u}\Delta(i,\bar{i}^{d})] \big\arrowvert_{u=u(i-1,\bar{i}^{d})}\\
\leq & [p_{u}U_{u}-\alpha p_{u}\Delta(i+1,\bar{i}^{d})] \big\arrowvert_{u=u(i,\bar{i}^{d})}\\
& -[p_{u}U_{u}-\alpha p_{u}\Delta(i,\bar{i}^{d})] \big\arrowvert_{u=u(i,\bar{i}^{d})}\\
= & -\alpha p_{u(i,\bar{i}^{d})}(\Delta(i+1,\bar{i}^{d})-\Delta(i,\bar{i}^{d})).
\end{array}
\end{equation}
The inequality holds because $\phi(\Delta(i,\bar{i}^{d}))$ is evaluated at $u=u(i,\bar{i}^{d})$ rather than at the optimal policy $u=u(i-1,\bar{i}^{d})$, and therefore it results in a higher than the optimal cost which yields an upper bound.

(2) Similarly 
\begin{equation}
\begin{array}{ll}
& \phi(\Delta(i+1,\bar{i}^{d}))-\phi(\Delta(i,\bar{i}^{d}))\\
= & [p_{u}U_{u}-\alpha p_{u}\Delta(i+1,\bar{i}^{d})] \big\arrowvert_{u=u(i,\bar{i}^{d})}\\
& -[p_{u}U_{u}-\alpha p_{u}\Delta(i,\bar{i}^{d})] \big\arrowvert_{u=u(i-1,\bar{i}^{d})}\\
\geq & [p_{u}U_{u}-\alpha p_{u}\Delta(i+1,\bar{i}^{d})] \big\arrowvert_{u=u(i-1,\bar{i}^{d})}\\
& -[p_{u}U_{u}-\alpha p_{u}\Delta(i,\bar{i}^{d})] \big\arrowvert_{u=u(i-1,\bar{i}^{d})}\\
= & -\alpha p_{u(i-1,\bar{i}^{d})}(\Delta(i+1,\bar{i}^{d})-\Delta(i,\bar{i}^{d})).
\end{array}
\end{equation}
This inequality holds also by a similar argument. \hfill\(\Box\)

Lemma 1 and Lemma 2 provide the monotonicity property as well as bounds on $\phi(\cdot)$ function differences between adjacent states. We next use these bounds to prove three monotonicity properties of $\Delta(i,\bar{i}^{d})$ with respect to state space parameter changes in RSR signal value $y$, current aggregated demand $i$, and tracking error $e$. Properties of $\Delta(i,\bar{i}^{d})$ will be used to prove the main Theorem on the structure of the optimal policy at the end of the section.

\subsection{Monotonicity of $\Delta(i,\bar{i}^{d(u)})$ for Key State Space Parameters}
We first discuss the monotonicity of $\Delta(i,\bar{i}^{d})$ for a fixed ISO RSR signal $y$. In this case $\bar{i}^{d}$ will be fixed and $i$, $e$ will change in the same direction. $\Delta(i,\bar{i}^{d})$ represents the optimal value difference between two adjacent states having only one consumption difference. Proposition 3 provides properties of $\Delta(i+1,\bar{i}^{d})$ when state space variable $\bar{i}^{d}$ is fixed while $i$ varies.

\textbf{Proposition 3} The following properties hold for a fixed $y$:

(1) $\Delta(i+1,\bar{i}^{d}) \geq \Delta(i,\bar{i}^{d}) + \epsilon^{l}$, where  $\epsilon^{l} = \frac{2\kappa}{1-\alpha[1-2(\lambda+\mu)-\upsilon]}$ with $\upsilon= \lambda(N-N_{1})$.

(2) $\Delta(i,\bar{i}^{d}) + \epsilon^{u} \geq \Delta(i+1,\bar{i}^{d})$, where $\epsilon^{u} = \frac{2\kappa}{1-\alpha[1-2(\lambda+\mu)]}$.

\textit{Proof.} See Appendix \ref{App:AppendixA}.

\textbf{Remark 4} Since both $\epsilon^{l}$ and $\epsilon^{u}$ are positive, we have $\Delta(i+1,\bar{i}^{d}) > \Delta(i,\bar{i}^{d})$. The optimal value function $J(i,\bar{i}^{d})$ exhibits convex-like behavior for a given $\bar{i}^{d}$, in the sense that 
\begin{equation}
J(i+1,\bar{i}^{d}) + J(i-1,\bar{i}^{d}) > 2J(i,\bar{i}^{d}).
\end{equation}
This convexity property can be used to design approximate DP (ADP) algorithms with convex functional approximation. We explore this possibility in Sec.~\ref{approximate dp algorithm}.

We next discuss the monotonicity of $\Delta(i,\bar{i}^{d})$ for a fixed tracking error $e$ with $\bar{i}^{d}$ and $i$ changing accordingly. Since one of our objectives is to accurately track the ISO RSR signal, it is reasonable to speculate that the SBO would use the same optimal policy for states $\{i,\bar{i}^{d}\}$ and $\{i+1,\bar{i}^{d}+\Delta y \}$ that have the same $e$, and therefore it is reasonable to have $\Delta(i+1,\bar{i}^{d}+\Delta y)=\Delta(i+1,\bar{i}^{d}+\Delta y)$. However, this speculation ignores the fact that the expected consumer arrival rate and the expected period utility reward will be different if same policy is used since the queuing system is closed and the total number of appliances is finite (different current consumption $i$ infers different number of counterpart idle appliances). We formally investigate the properties of $\Delta(i,\bar{i}^{d})$ for a given $e$ and state the monotonicity properties as follows.

\textbf{Proposition 4} The following properties hold for a fixed $e$:

(1) $\Delta(i,\bar{i}^{d})\geq\Delta(i+1,\bar{i}^{d}+\Delta y)$.

(2) $\Delta(i+1,\bar{i}^{d}+\Delta y)+\bar{\epsilon}^{u} \geq\Delta(i,\bar{i}^{d})$ where $\bar{\epsilon}^{u} = \frac{\alpha(\lambda+\mu)}{1-\alpha[1-(\lambda+\mu)]}\epsilon^{u}$.


The proof of Proposition 4 is similar to the proof of Proposition 3. We omit the proof due to page limitation. 

In the end, we derive a last property when the aggregated consumption $i$ is fixed while the ISO RSR signal $y$ and tracking error $e=i-\bar{i}-yR$ change in the same direction. 

\textbf{Proposition 5} The following properties hold for a fixed $i$:

(1) $\Delta(i,\bar{i}^{d}) \geq \Delta(i,\bar{i}^{d}+\Delta y) + \epsilon^{l}$.

(2) $\Delta(i,\bar{i}^{d}+\Delta y) + \epsilon^{u} + \bar{\epsilon}^{u} \geq \Delta(i,\bar{i}^{d})$.

\textit{Proof.} We have 
\begin{equation}
\Delta(i,\bar{i}^{d}) \geq \Delta(i+1,\bar{i}^{d}+\Delta y) \geq \Delta(i,\bar{i}^{d}+\Delta y) + \epsilon^{l},
\end{equation}
where the first (second) inequality is the direct result of Proposition 4 (3). And similarly the second part of the proposition holds. \hfill\(\Box\)

The above proposition completes our discussion of the properties of the differential cost function $\Delta(i,\bar{i}^{d})$. These properties and the relation between $\Delta(i,\bar{i}^{d})$ and $u(i,\bar{i}^{d})$ result in the following useful properties of the optimal policy in the next section.

\subsection{Monotonicity Properties of the Optimal Policy $u(i,\bar{i}^{d})$}
Based on Propositions 3, 4, and 5, we present the following theorem as the main result illustrating the monotonicity properties of the optimal policy $u(i,\bar{i}^{d})$.

\textbf{Theorem 1} The following properties hold for the state feedback optimal policy $u(i,\bar{i}^{d})$ for all $\{i,\bar{i}^{d}\}$:

(1) For the same RSR signal, the optimal price policy is a monotonically non-decreasing function of $i$. Namely $u(i+1,\bar{i}^{d})\geq u(i,\bar{i}^{d})$ .

(2) For the same tracking error, the optimal price policy is a monotonically non-increasing function of $i$. Namely $u(i,\bar{i}^{d})\geq u(i+1,\bar{i}^{d}+\Delta y)$.

(3) For the same consumption level, the optimal price policy is a monotonically non-increasing function of $\bar{i}^{d}$. Namely $u(i,\bar{i}^{d})\geq u(i,\bar{i}^{d}+\Delta y)$.

\textit{Proof.} The proof is straightforward. From Proposition 3, 4, and 5 we have
\begin{displaymath}
\Delta(i,\bar{i}^{d})\geq\Delta(i+1,\bar{i}^{d}+\Delta y)\geq\Delta(i,\bar{i}^{d}+\Delta y).
\end{displaymath}
From Proposition 2 the optimal control is a non-decreasing function of the $\Delta(i,\bar{i}^{d})$, therefore
\begin{displaymath}
u(i,\bar{i}^{d})\geq u(i+1,\bar{i}^{d}+\Delta y)\geq u(i,\bar{i}^{d}+\Delta y),
\end{displaymath}
and the three statements above are true. \hfill\(\Box\)

\textbf{Remark 5} The policy monotonicity properties are valid for both $D(t)=1$ and $D(t)=-1$. Bang-bang optimal control will be used when greater imbalance exists between the aggregated consumption and ISO regulation signal level. When the state is at a high value of $i$ and a low value of $\bar{i}^{d}$, the ISO would broadcast highest price signal. When the state has a small value of $i$ and a large $\bar{i}^{d}$, the SBO would broadcast a lower price signal. Otherwise, the SBO would broadcast the optimal price in between.

%
%
%

\textbf{Remark 6} The three monotonicity properties can be interpreted as follows: we set a low price for states with a higher RSR signal, and a high price for states with a higher aggregate consumption. When the tracking error is the same, a high current consumption suggests a smaller number of disconnected appliances, which means we have a smaller number of appliances that are able to respond to the SBO's signal. In such cases, a lower price is chosen to achieve a larger percentage of disconnected appliances.

\subsection{Monotonicity of Price Sensitivity w.r.t. $N$}\label{price sensitivity}
We investigate the change in the optimal price sensitivity with respect to the number of connected appliances $i$, as the total number of appliances $N$ and the maximum Reserve obligation $R$ increase both in a constant proportion. Consider a SBO who provides regulation reserves $R$ equal to a fixed proportion $q$ of the duty cycle appliances $N$ \footnote{Using the average consumption rate of an active appliance as the unit that measures capacity, $N$ equals the maximal consumption rate of the cooling appliances under the SBO's control.}, namely $R=qN$. Note that $q$ can be determined by cooling appliance user preferences, appliance technical specifications, etc \cite{BowenBaillieul2013}. As $N$ and $R$ increase at the same rate, the effective penalty parameter $\kappa=K/R^{2}$ will decrease. In addition, the uniformized policy update interval $\Delta_{t}$ and discount factor $\alpha$ will change as $N$ increases following the relationships:
 \begin{equation}\label{dt with N}
\Delta_{t} = \frac{1}{N\max(\lambda,\mu)+\gamma}\approx\frac{1}{N\max(\lambda,\mu)},
\end{equation}
and 
\begin{equation}\label{alpha with dt}
\alpha = \frac{1}{1+r\Delta_{t}}
\end{equation}
where $r$ is the prevailing discount rate.
Substituting (\ref{dt with N}) into (\ref{alpha with dt}) we can write
\begin{equation}
\alpha = \frac{N\max(\lambda,\mu)}{r+N\max(\lambda,\mu)}.
\end{equation}
Observing that both the discount rate and the policy update period increase as $N$ increases, we show by Proposition 6 that the change in the value function differentials 
$\Delta(i+1,\bar{i}^{d}) - \Delta(i,\bar{i}^{d})$ and 
$\Delta(i,\bar{i}^{d}) - \Delta(i,\bar{i}^{d}+\Delta y)$ approaches zero as $N$ approaches infinity.

\textbf{Proposition 6} $\epsilon^{u}$ and $\epsilon^{l}$, defined in Proposition 3, and $\bar{\epsilon}^{u}$, defined in Proposition 4,  will decrease as $N$ increases, and moreover for $N\rightarrow\infty$, their asymptotic limit is $0$.

\textit{Proof.} Using explicitly $\Delta_{t}$ which for notational simplicity was selected as the time unit and set equal to $1$ in the proof of Proposition 3, we can write 
\begin{equation}\label{eq46}
\epsilon^{u}=\frac{2\kappa\Delta_{t}}{1-\alpha[1-2(\lambda+\mu)\Delta_{t}]}.
\end{equation}
Substituting into (\ref{eq46}) the effective discount factor $\alpha$ and the relation $\Delta_{t}\approx 1/(N\max(\lambda+\mu))$ we obtain
\begin{equation}
\epsilon^{u}(N) = \frac{\frac{2K/(qN)^{2}}{N\max(\lambda,\mu)}}{1-\frac{N\max(\lambda,\mu)}{r+N\max(\lambda,\mu)}[1-2(\lambda+\mu)\frac{1}{N\max(\lambda,\mu)}]}
\end{equation}
which in turn simplifies to $\epsilon^{u}(N) = \frac{2K/(qN)^{2}}{r+2(\lambda+\mu)}$ verifying that $\epsilon^{u}$ decreases as $N$ increases. It can be similarly shown that $\epsilon^{l}$ also decreases as $N$ increases. Finally $\bar{\epsilon}^{u}$ is shown below to equal a positive multiple of $\epsilon^{u}$
\begin{equation}
\begin{array}{lll}
\bar{\epsilon}^{u} & = & \frac{\alpha(\lambda+\mu)\Delta_{t}}{1-\alpha[1-(\lambda+\mu)\Delta_{t}]}\epsilon^{u},\\
& = & \frac{\frac{N\max(\lambda,\mu)}{r+N\max(\lambda,\mu)}\Delta_{t}}{1-\frac{N\max(\lambda,\mu)}{r+N\max(\lambda,\mu)}[1-\frac{\lambda+\mu}{N\max(\lambda,\mu)}]}\epsilon^{u},\\
& = & \frac{1}{r+\lambda+\mu}\epsilon^{u}.
\end{array}
\end{equation}
We can now conclude that all three parameters $\bar{\epsilon}^{u}$, $\epsilon^{u}$ and $\epsilon^{l}$ will approach zero as $N$ goes to infinity. \hfill\(\Box\)

Proposition 6 describes the asymptotic impact of building size described by $N$ on $\Delta(i,\bar{i}^{d})$ and the optimal price policy $u(i,\bar{i}^{d})$. According to Propositions 3 and 5, the difference between differential cost functions for fixed $\bar{i}^{d}$ and $i$ respectively is bounded by
\begin{equation}
\Delta(i+1,\bar{i}^{d}) - \Delta(i,\bar{i}^{d}) \in [\epsilon^{l},\epsilon^{u}]
\end{equation}
and
\begin{equation}
\Delta(i,\bar{i}^{d}) - \Delta(i,\bar{i}^{d}+\Delta y) \in [\epsilon^{l},\epsilon^{u} + \bar{\epsilon}^{u}].
\end{equation}
which by Proposition 6 implies that these differences go to $0$. 

Using the expression for the optimal policy proven in Proposition 2, we can conclude that $u(i,\bar{i}^{d})$, $u(i+1,\bar{i}^{d})$,  $u(i,\bar{i}^{d})$ and $u(i,\bar{i}^{d}+\Delta y)$ get closer together as $N$ increases, and as a result the optimal policy function becomes flatter with respect to its arguments.

\section{Numerical Solution Algorithms}
\label{approximate dp algorithm}

The analytical results presented so far are not merely exercises in analysis that capture abstract properties of the DP optimality conditions. Most notably, the optimal policy structure of Proposition 2 and the monotonicity and second derivative related properties proven in Propositions 3 to 5 are valuable ammunition that enable design and implementation of efficient and scalable numerical solution algorithms. This section demonstrates the value of the analytical results in doing just that and provides elaborative computational results. 

\subsection{Value Iteration Based Approaches}
We first propose and implement two numerical DP solution algorithms, the first for benchmarking and comparison purposes using the conventional value iteration (CVI) approach \cite{Bertsekas2007}, and the second by leveraging the optimal policy structure proven in Proposition 2 of Section IV which we call \textit{assisted value iteration} (AVI) algorithm. The AVI algorithm replaces the computationally inefficient discretization of the allowable policy space and exhaustive search over it at each iteration. We instead use the policy in (\ref{saturation control}) because it is optimal for a given value function resembling policy iteration algorithms. Our AVI algorithm recognizes that the state space is discrete while the policy space is continuous. It benefits from (1) the analytic characterization of the optimal policy in terms of the current iteration estimate of the value function thus avoiding both state space discretization and exhaustive search for the optimal policy, and (2) avoidance of the sub-optimality gap introduced by the policy space discretization. 

\begin{figure}[hbt]
\centering

\includegraphics[width=9cm]{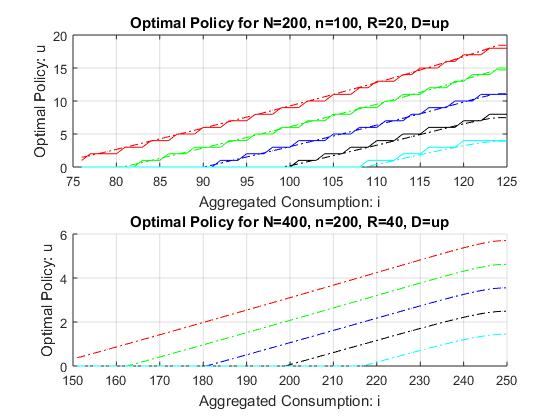}

\caption{Optimal policies obtained from the CVI and AVI Algorithms. The solid (dashed) lines are simulation results of CVI (AVI) algorithm. The red, green, dark black, black, and light black curves correspond to RSR signal level at -0.8, -0.4, 0.0, 0.4, 0.8, respectively. In CVI, the policy space is discretized into 20 possible prices corresponding to the temperature threshold from 1 to 20. In AVI, the policy space is continuous. Monotonicity properties in Theorem 1 are verified. Equal $Y=yR$ policy functions demonstrate that the slope of the price between $u=T_{\min}$ to $u=T_{\max}$ decreases as $N$ and $R$ increase. The vertical distance among price policy lines also decreases.}

\label{compare_asymptotic_results}
\end{figure}

Numerical results from the CVI and AVI algorithms are shown in Fig.~\ref{compare_asymptotic_results}. In the upper sub-figure, the parameter values used were $N=200$, $\bar{n}=100$, $R=20, \lambda=2, \mu=0.5$. We choose a linear utility function as in (\ref{utility function}) with $b=20$. We find that the CVI algorithm yields policies selected from the discretized allowable policy set and the AVI algorithm provides a smooth and continuous policy. The observed price monotonicity are consistent with properties derived in Theorem 1. The comparison between the two sub-figures demonstrates the price sensitivity when we increase $N$ and $R$ to the same proportion. Note that the rate at which the optimal price increases from $u=T_{\min}$ to $u=T_{\max}$ decreases, unsurprisingly, by a factor of 2. Another interesting observation is that when $N$ increases, different curves for ISO signals $y$ get closer to each other for a fixed $i$. This is consistent with our analysis of the monotonicity of price sensitivity in Sec. \ref{price sensitivity}.

\subsection{Functional Approximate DP Approach}
We proceed to propose a numerical solution algorithm based on an analytic functional approximation of the value function $J(i, y, D)$. This algorithm leverages the properties of value function first and second differences derived in Sec.~\ref{section:Optimal dynamic policies}. In particular we use Proposition 3 which shows that $\Delta(i+1,\bar{i}^{d})\geq \Delta(i,\bar{i}^{d})$. Given the discrete state space of our problem, this property is equivalent to convexity of $J(i,\bar{i}^{d})$ in the number of active appliances $i$ for a given pair of ISO signal's value and direction. 

In addition, we note that from Fig.~\ref{compare_asymptotic_results} that the increased rate of the optimal policy for a fixed RSR signal is approximately a constant value $k$. Therefore we approximately have
\begin{equation}
u(i+1,\bar{i}^{d})-u(i+1,\bar{i}^{d})=k,
\end{equation}
which is equivalent to have $\Delta(i+1,\bar{i}^{d})-\Delta(i,\bar{i}^{d})$ being some constant. Since $\Delta(i,\bar{i}^{d})$ is the differential of the value function with respect to $i$, it means that the second order differential of the value function with respect to $i$ is approximately constant, namely $\frac{\partial^{2}}{\partial i^{2}}J(i,y,D)$ is approximately constant. Similarly, from Fig.~\ref{compare_asymptotic_results} we note that the rate of policy's vertically changes is constant for varying RSR signals, therefore the $\frac{\partial}{\partial \bar{i}^{d}}\Delta(i,\bar{i}^{d})$ is approximately constant, namely $\frac{\partial}{\partial \bar{i}^{d}}\frac{\partial}{\partial i}J(i,\bar{i}^{d})$ is constant.

These properties motivate an approximation of $J(i,\bar{i}^{d})$ by $\hat{J}_{d}(i,y)$ that is quadratic in $i-yR$. In fact, we treat $D(t)$ as a binary argument and  propose the following basis function approximation when $D(t) = -1$.
\begin{equation}\label{lower freedom approximation}
\hat{J}_{d}(i,y) = r_{11}(i-yR)^2 + r_{12}(i-yR) + r_{13}
\end{equation}
with $r_{11} > 0$ to guarantee convexity. In addition, the differential of the value function with respect to $i$, namely $\Delta(i,\bar{i})$, is 
\begin{equation}
\label{ea53}
\frac{\partial }{\partial i}\hat{J}_{d}(i,y) = \Delta(i,\bar{i}) = 2r_{11}(i-yR) + r_{12}i.
\end{equation}
Proposition 4 proved that $\Delta(i,\bar{i}^{d})\geq \Delta(i+1,\bar{i}^{d}+\Delta y)$ which suggests that $\frac{\partial }{\partial i}\hat{J}_{d}(i,y)$ monotonically decreases as a function of $i$ for a fixed $e=i-yR$, hence $r_{12} < 0$ in (\ref{ea53}).

We generalize the approximation in (\ref{lower freedom approximation}) by differentiating the parameters depending on the discrete value of the direction $D$, Thus we define for $D = -1$
\begin{equation}\label{feature parameters}
\hat{J}_{d}(i,y,\mathbf{r_{d}}) = r_{1}i^2 + r_{2}i + r_{3}y^2 + r_{4}y + r_{5}iy + r_{6}
\end{equation}
and the corresponding function $\hat{J}_{u}(i,y,\mathbf{r_{u}})$ for $D = 1$. The value function $J(i,y,D,\mathbf{r})$ is then approximated by:
\begin{equation}
\label{value function approximation}
\hat{J}(i,y,D,\mathbf{r})  =  \mathbbm{1}_{\{D = 1\}} \hat{J}_{u}(i,y,\mathbf{r_{u}})+ \mathbbm{1}_{\{D = -1\}} \hat{J}_{d}(i,y,\mathbf{r_{d}}).
\end{equation}
The two components of the expression in (\ref{value function approximation}) approximate state features associated with increasing or decreasing ISO signals. The vector $\textbf{r}$ is a vector of twelve parameters six from $\mathbf{r_{d}}$ and $\mathbf{r_{u}}$ each. Written in matrix form, (\ref{value function approximation}) is equivalent to the following
\begin{equation}
\hat{J} = \Phi \mathbf{r},
\end{equation}
where $\Phi$ is a $|N|\times|y|\times|D|$ by 12 matrix with rows being the feature vector for each state. The functional approximation is therefore transformed into the problem of solving the projected Bellman equation
\begin{equation}
\Phi \mathbf{r} = \Pi T(\Phi \mathbf{r}),
\end{equation}
where $T$ is the operator of the form $TJ = g+\alpha PJ$ with state transition matrix $P$. $\Pi$ is the projection operator onto the set spanned by the basis functions $S=\{\Phi \mathbf{r}| \mathbf{r}\in \Re^{12}\}$. It is shown in \cite{Bertsekas2007} that the solution to the above projection problem is given by
\begin{equation}
\label{optimal parameter r}
\mathbf{r}^{\star} = C^{-1}d,
\end{equation}
where $C=\Phi^{'}\Xi(I-\alpha P)\Phi$, $d=\Phi^{'}\Xi g$, and $\Xi$ is the matrix with diagonal elements being the steady state probability distribution of the states. It is further shown that (\ref{optimal parameter r}) can be solved in an iterative form by the projected value iteration (PVI) algorithm
\begin{equation}
\label{eq59}
\mathbf{r_{k+1}} =\mathbf{r_{k}} - \gamma G_{k}(C_{k}\mathbf{r_{k}}-d_{k}),
\end{equation} 
where $C_{k}$ and $d_{k}$ are given by
\begin{equation}
\label{eq60}
\begin{array}{c}
C_{k}=\frac{1}{k+1}\sum\limits_{t=0}^{k}\phi(i_{t},y_{t},D_{t})(\phi(i_{t},y_{t},D_{t})-\alpha\phi(i_{t+1},y_{t+1},D_{t+1})),\\
d_{k}=\frac{1}{k+1}\sum\limits_{t=0}^{k}\phi(i_{t},y_{t},D_{t})g(i_{t},y_{t},D_{t},u(t)).
\end{array}
\end{equation}
To choose $\gamma$ and $G_{k}$, it is proposed to have $\gamma=1$ and 
\begin{equation}
\label{eq62}
G_{k}=(\frac{1}{k+1}\sum\limits_{t=0}^{k}\phi(i_{t},y_{t},D_{t})\phi(i_{t},y_{t},D_{t})^{'})^{-1}.
\end{equation}
Based on the above approach that finds a good approximation of the value function for a \textit{fixed} policy, we successfully use the following algorithm to construct a good approximation of the value function as well as the \textit{optimal} policy based on sample trajectories obtained by Monte Carlo simulation. The algorithm contains the following four steps:

\vspace{5mm}

\noindent\textbf{ADP Algorithm}
\vspace{-3mm}

\noindent\makebox[\linewidth]{\rule{\columnwidth}{0.4pt}}
\textbf{Step 1.} \textbf{Initialization} \textbf{r}=0.

\textbf{Step 2.} \textbf{Initialization} $\mathbf{r}_{old}=\mathbf{r}$, $\mathbf{r}_{0}=\mathbf{r}$, $k=0$, $\{i_{k},y_{k},D_{k}\}$.

\textbf{Step 3.} \textbf{Generate} 

\hspace{15mm} Optimal policy $u_{\mathbf{r_{old}}}(i_{k},y_{k},D_{k})$ 

\hspace{15mm} Next state $\{i_{k+1},y_{k+1},D_{k+1}\}$ 

\hspace{15mm} Period cost $g(i_{k},y_{k},D_{k},u_{\mathbf{r_{old}}}(i_{k},y_{k},D_{k}))$

\hspace{12mm}\textbf{Update} 

\hspace{15mm} $C_{k}, d_{k}$, and $\mathbf{r}_{k+1}$ based on (\ref{eq59})--(\ref{eq62})

\hspace{11mm} \textbf{If} $k\geq k_{\min}$ and $||\mathbf{r}_{k+1}-\mathbf{r}_{k}||_{2}<\bar{\epsilon}$

\hspace{15mm} $\mathbf{r}=\mathbf{r}_{k+1}$, go to \textbf{Step 4}.

\hspace{11mm} \textbf{Else} 

\hspace{15mm} $k=k+1$, go to \textbf{Step 3}.

\textbf{Step 4.} \textbf{If} $||J(i,y,D,\mathbf{r})-J(i,y,D,\mathbf{r_{old}})||_{\infty}<\tau$

\hspace{15mm} return $\mathbf{r^{\star}}=\mathbf{r}$. Algorithm ends.

\hspace{11mm} \textbf{Else} go to \textbf{Step 2}.
\vspace{-3mm}

\noindent\makebox[\linewidth]{\rule{\columnwidth}{0.4pt}}
\vspace{1mm}

The algorithm starts with an initial guess of the parameters, $\mathbf{r}=0$. Step 2 initializes the iteration count, parameters $\mathbf{r}_{0}$, and state variables $\{i_{0},y_{0},D_{0}\}$. Step 3 iteratively updates the value function for the fixed policy $u_{\mathbf{r_{old}}}$ using the PVI algorithm described above. Step 3 is repeated for at least a minimum number of iterations, $k\geq k_{\min}$, and stops when the chage in $\mathbf{r}$ meets a desired tolerance, $||\mathbf{r}_{k+1}-\mathbf{r}_{k}||_{2}<\bar{\epsilon}$. Steps 4 compares the value function parametrized by $\mathbf{r_{old}}$ and the updated \textbf{r} obtained by step 3. If the infinity norm of the vector is less than the threshold $\tau$, then the value function converges and the algorithm returns the optimal parameter $\mathbf{r^{\star}}=\mathbf{r}$. Else, it returns to step 2 for a new iteration.

\subsection{Comparison between Value Iteration Algorithms and the ADP}
We compare the computational performance of the CVI, AVI and Functional ADP algorithms for different state space size problems in Table I. Based on the optimal condition derived in Proposition 2, the AVI algorithm is effective in reducing the computational time by approximately 90\% since the optimal policy per state and per iteration is calculated on the fly based on the current value function. However, it is not fast for large problems up to 400,000 states since it needs more than 2 hours to solve for the optimal policy. Considering that the RSR is bid and served for every one hour, the AVI algorithm may not be practical for real time implementation, especially when the RSR provision capacity of the energy provider is huge. However, the functional approximation based ADP algorithm further reduces the computational time by more than 90\% from the AVI algorithm. In fact, in the inner loop described in Algorithm 1, the number of states visited in Monte Carlo simulation is approximately 10\% of all the states. As for the outer loop, it also needs few iterations for the convergence of the value function compared to the AVI algorithm. Therefore the total computational time is reduced for more than 90\%.

\begin{table}[h]
\label{table1}
\caption{computational performances of the three algorithms}
\centering
\begin{tabular}{|c|c|c|c|}
\hline Problem Size ($|N|*|y|*|D|$) & 50*21*2 & 500*41*2 & 5000*41*2 \\ 
\hline  CVI Computation Time (sec)& 200.35 & 6761.2 & 66082 \\ 
\hline  AVI Computation Time (sec)& 20.14  & 596.65  &  9489.5\\ 
\hline  ADP Computation Time (sec)& 2.69 & 34.2 & 514.8 \\ 
\hline 
\end{tabular} 
\end{table}

\begin{figure}[h]
	\centering
	\includegraphics[width=9cm]{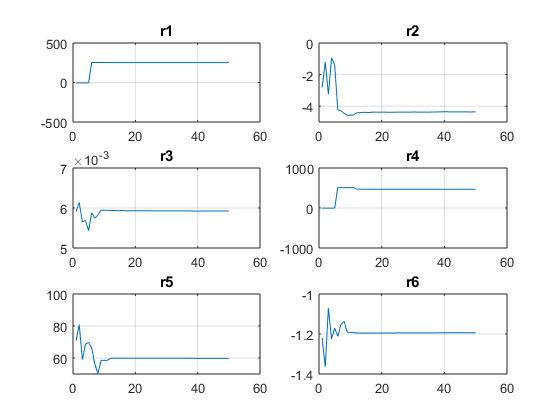}
		\caption{Convergence of the feature parameters in (\ref{feature parameters}). The coefficients of the six basis converge after 10 iterations. Convergence of the feature parameters indicates the convergence of the value function approximation $\hat{J}$.}
	\label{convergence of parameter}
\end{figure}

We examine the convergence result of the proposed algorithm. The coefficient vector $\mathbf{r}$ is composed of 12 parameters, six for each direction $D=1,-1$.  Fig.~\ref{convergence of parameter} shows the convergence result of the six parameters for direction $D=-1$ proposed by the ADP algorithm. We find all parameters would converge after 10 iterations. The convergence of $\mathbf{r}$ also indicates the convergence of the value function $\hat{J}$.

\begin{figure}[h]
	\centering
	\includegraphics[width=9cm]{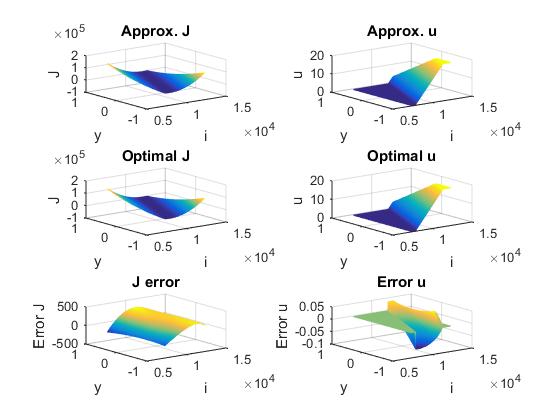}
		\caption{Left column figures compare at the top the value functions obtained by the functional approximation DP algorithm and the AVI algorithm, and plot the error at the bottom. Right column figures plot the corresponding optimal policy and the error amongst them. Results indicates that the functional approximation DP algorithm captures the properties proven in section IV.}
	\label{critic_only_value_and_ctrl_big_space}
\end{figure}

Fig.~\ref{critic_only_value_and_ctrl_big_space} compares the converged value function and the corresponding optimal policy generated by the AVI and the functional ADP. Left column figures are plots of the value function generated by the AVI and the functional ADP algorithm corresponding to ISO RSR signal direction $D = 1$. The functional ADP algorithm learns the convex structure of the value function accurately. The error between $\hat{J}(i,y,D,\mathbf{r})$ and $J(i,y,D)$ is relatively small. Right column figures compare the optimal policies of Proposition 2 that are generated by the value functions based on the functional ADP and the AVI algorithm. The functional ADP algorithm performs well relative to the AVI algorithm that derives the true optimal policy.

{\color{black}
\subsection{RSR Provision Quality with Dynamic Preference}
We validate  the tracking performance of the ADP algorithm optimal control with RSR signals from PJM. To illustrate the improvement gained by having dynamic preference, we compare results to RSR provision where the optimal policy is generated based on the assumption of {\it time invariant} consumers' preference, namely to assume that consumers' preferences distribution is statically uniformly distributed regardless of time. In this simulation we assume that there are 20,000 appliances within the aggregator each consuming 1kW of energy. The aggregator has mean consumption of 10,000 kW and provide 2,000 kW up (down) reserve to the market, namely 10\% of its total capacity. Fig.~\ref{tracking quality comparison} shows the comparison of the tracking performance between the two assumptions. The red curve is the reference PJM signal at a frequency of 4 seconds. The black curves are the tracking performance (upper figure) and the tracking error (lower figure) that result from the dynamic preferences optimal control. The green curves are results from the static preferences optimal control. In general, both approaches are effective in providing demand side RSR with tracking error controlled within 5\%. Nonetheless, the tracking provided by dynamic preference assumption achieves higher accuracy. The accuracy can be seen based on the statistics in Table II. The mean tracking error is decreased from 86MW to 79MW if dynamic preference is applied, namely a mean error reduction of 8\%. The min and max tracking error spikes are also decrease by 12\% and 17\%, respectively. As to the aggregator, the sum of squared tracking penalty is decreased significantly by 9.4\%. 

Converting tracking error to the system operation perspective, with more accurate dynamic preference modeling, the system will have 8\% less total reserve generation and 17\% less peak reserve generation from traditional thermal units. These will provide significant operation values since these capacity can otherwise be committed as energy, rather than reserve, and thus possibly lower the overall energy costs across the system either in the day ahead or real time market clearing.

\begin{figure}[hbt]
	\centering
	\includegraphics[width=9cm]{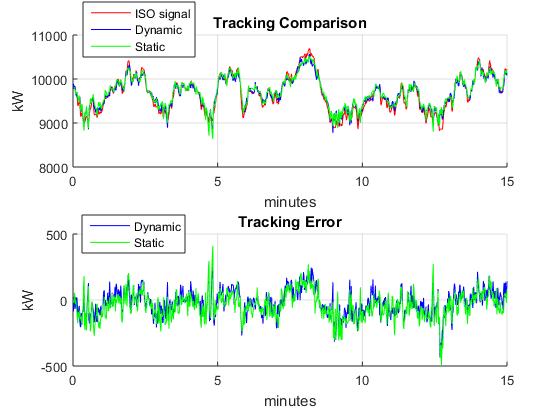}
	\caption{Comparison of the optimal control generated based on the assumption of dynamic preference vs. static uniform preference. In both assumptions, the optimal control can track the trend of RSR signal. Dynamic preferences optimal control tracks with better statistical performances of the tracking error (mean, standard deviation, max, and min) than the uniform preference optimal control.}
	\label{tracking quality comparison}
\end{figure}

\begin{table}[h]
	\label{table2}
	\caption{Dynamic vs. Static Preference RSR Provision Quality}
	\centering
	\begin{tabular}{|c|c|c|c|}
		\hline Statistics & Dynamic Preference & Static Preference\\ 
		\hline Mean Error (kW)& 79.18 & 86.26 \\
		\hline Std.Dev. Error (kW)& 62.31 & 68.25 \\
		\hline Min Error (kW)& -437 & -492 \\
		\hline Max Error (kW)& 342 & 410 \\
		\hline Sum of Squared Penalty & 913330 & 1008400 \\
		\hline 
	\end{tabular} 
\end{table}
}
\section{Conclusion}
\label{conclusion}
This paper relaxes a common, but unrealistic assumption in the dynamic pricing literature, which, for the sake of simplifying the analysis of the resulting problem formulation, claims that it is reasonable to approximate the preferences of market participants with a static, usually uniform, distribution that is independent of control history. We show that a dynamically evolving trapezoidal pdf captures the dynamics of market participant preferences in the cooling appliance duty cycle paradigm considered here, proceed to model dynamic preferences, and succeed to overcome the complexity that it introduces. We believe that dynamic, control driven evolution of preferences can be modeled and analyzed in more general contexts. Under preference dynamics modeling, we derive an analytic expressions of the optimal policy, monotonicity properties of the value function, and the behavior of the optimal policy. We also prove the existence of policy sensitivity bounds and their asymptotic convergence as the number of the duty cycle appliances increase. The aforementioned analytical results prove invaluable in guiding us to design and implement efficient and scalable numerical solution algorithms. {\color{black} Simulation based on PJM signals shows the advantage of RSR provision under dynamic preference modeling over static modeling assumptions. In future work we will consider to prove the convergence of the ADP algorithm, to embed additional sufficient statistics into the dynamic preference, and to model imperfect state observation where the utility function or the preferences cannot be observed.}

\appendices
\section{Proof of Proposition 3} \label{App:AppendixA}
\textit{Proof.} (1) For a sequence of $J_{0}(i,\bar{i}^{d}),\ldots,J_{k}(i,\bar{i}^{d})$ generated by value iteration, we have $\lim\limits_{k\rightarrow\infty}J_{k}(i,\bar{i}^{d})=J(i,\bar{i}^{d})$ based on the value iteration convergence property. We define the differential of the value function at the $k^{\mathrm{th}}$ iteration as
\begin{displaymath}
\Delta_{k}(i,\bar{i}^{d})=J_{k}(i,\bar{i}^{d})-J_{k}(i-1,\bar{i}^{d}).
\end{displaymath}
It follows that $\lim\limits_{k\rightarrow\infty}\Delta_{k}(i,\bar{i}^{d})=\Delta(i,\bar{i}^{d})$. 

We assume $\Delta_{k}(i+1,\bar{i}^{d})\geq\Delta_{k}(i,\bar{i}^{d})+\epsilon_{k}^{l}$ and $\Delta_{k}(i+1,\bar{i}^{u})\geq\Delta_{k}(i,\bar{i}^{u})+\epsilon_{k}^{l}$ with $\epsilon_{k}^{l}=0$ for all $\{i,\bar{i}^{d}\}$, $\{i,\bar{i}^{u}\}$ and $k=0$, which holds trivially when at the initial iteration the value function is taken to equal zero. At iteration $k+1$, $J_{k+1}(i,\bar{i}^{d})$ can be written using the Bellman equation as
\begin{equation}
\label{value function}
\begin{array}{lll}
J_{k+1}(i,\bar{i}^{d})&=& g(i,\bar{i}^{d})-\lambda(N-i)\phi(\Delta(i+1,\bar{i}^{d}))+\\
&& \alpha\big\{[1-(\gamma_{1}^{d}+\gamma_{2}^{d})]J_{k}(i,\bar{i}^{d}) -\mu i\Delta_{k}(i,\bar{i}^{d})\\
&& +(\gamma_{1}^{d}J_{k}(i,\bar{i}^{u}+\Delta y)+\gamma_{2}^{d}J_{k}(i,\bar{i}^{d}-\Delta y))\big\},
\end{array}
\end{equation}
Starting with the definition of $\Delta_{k}(i+1,\bar{i}^{d})$, we can write
\begin{displaymath}
\label{D xk+1}
\begin{array}{cll}
& \Delta_{k+1}(i+1,\bar{i}^{d})\\
= & J_{k+1}(i+1,\bar{i}^{d})-J_{k+1}(i,\bar{i}^{d}),\\
= & [g(i+1,\bar{i}^{d})-g(i,\bar{i}^{d})] + \alpha\big\{[1-\gamma_{1}^{d}-\gamma_{2}^{d}]\Delta_{k}(i+1,\bar{i}^{d})\\
& - \mu (i+1)\Delta_{k}(i+1,\bar{i}^{d})+\mu i\Delta_{k}(i,\bar{i}^{d})\\
& + \gamma_{1}^{d}\Delta_{k}(i+1,\bar{i}^{u}+\Delta y)+\gamma_{2}^{d}\Delta_{k}(i+1,\bar{i}^{d}-\Delta y)\big\}\\
& - \lambda(N-i-1)\phi(\Delta_{k}(i+2,\bar{i}^{d}))\\
& + \lambda(N-i)\phi(\Delta_{k}(i+1,\bar{i}^{d})),
\end{array}
\end{displaymath}
This can be used to derive the change in $\Delta_{k+1}(i,\bar{i}^{d})$ when $i$ increases by one,
\begin{equation}
\label{proposition 2 eq 1}
\begin{array}{ll}
&\Delta_{k+1}(i+1,\bar{i}^{d})-\Delta_{k+1}(i,\bar{i}^{d})\\
=& g(i+1,\bar{i}^{d})-2g(i,\bar{i}^{d})+g(i-1,\bar{i}^{d})\\
&+\alpha\big\{ (1-\gamma_{1}^{d}-\gamma_{2}^{d})[\Delta_{k}(i+1,\bar{i}^{d})-\Delta_{k}(i,\bar{i}^{d})]\\
& - \mu(i+1)[\Delta_{k}(i+1,\bar{i}^{d})-\Delta_{k}(i,\bar{i}^{d})]\\
& +\mu(i-1)[\Delta_{k}(i,\bar{i}^{d})-\Delta_{k}(i-1,\bar{i}^{d})]\\
& + \gamma_{1}^{d}[\Delta_{k}(i+1,\bar{i}^{u}+\Delta y)-\Delta_{k}(i,\bar{i}^{u}+\Delta y)]\\
& + \gamma_{2}^{d}[\Delta_{k}(i+1,\bar{i}^{d}-\Delta y)-\Delta_{k}(i,\bar{i}^{d}-\Delta y)]\big\}\\
& - \lambda (N-i-1)[\phi(\Delta_{k}(i+2,\bar{i}^{d}))-\phi(\Delta_{k}(i+1,\bar{i}^{d}))]\\
& + \lambda(N-i+1)[\phi(\Delta_{k}(i+1,\bar{i}^{d}))-\phi(\Delta_{k}(i,\bar{i}^{d}))].
\end{array}
\end{equation}
From Lemma 2, 
{\small
\begin{displaymath}
\phi(\Delta_{k}(i+2,\bar{i}^{d}))-\phi(\Delta_{k}(i+1,\bar{i}^{d})) \leq -\alpha p_{u(i+1,\bar{i}^{d})}[\Delta(i+2,\bar{i}^{d})-\Delta(i+1,\bar{i}^{d})],
\end{displaymath}
\begin{displaymath}
\phi(\Delta_{k}(i+1,\bar{i}^{d}))-\phi(\Delta_{k}(i,\bar{i}^{d})) \geq -\alpha p_{u(i-1,\bar{i}^{d})}[\Delta(i+1,\bar{i}^{d})-\Delta(i,\bar{i}^{d})].
\end{displaymath}}
We substitute the above two inequalities into (\ref{proposition 2 eq 1}),
\begin{equation}
\begin{array}{ll}
&\Delta_{k+1}(i+1,\bar{i}^{d})-\Delta_{k+1}(i,\bar{i}^{d})\\
\geq & g(i+1,\bar{i}^{d})-2g(i,\bar{i})+g(i-1,\bar{i}^{d})\\
&\alpha\{[1-\gamma_{1}^{d}-\gamma_{2}^{d}-\mu(i+1)][\Delta_{k}(i+1,\bar{i}^{d})-\Delta_{k}(i,\bar{i}^{d})]\\
& +\mu(i-1)[\Delta_{k}(i,\bar{i}^{d})-\Delta_{k}(i-1,\bar{i}^{d})]\\
& +\gamma_{1}^{d}[\Delta_{k}(i+1,\bar{i}^{u}+\Delta y)-\Delta_{k}(i,\bar{i}^{u}+\Delta y)]\\
& +\gamma_{2}^{d}[\Delta_{k}(i+1,\bar{i}^{d}-\Delta y)-\Delta_{k}(i,\bar{i}^{d}-\Delta y)]\}\\
& -\lambda(N-i+1)p_{u(i-1,\bar{i}^{d})}[\Delta_{k}(i+1,\bar{i}^{d})-\Delta_{k}(i,\bar{i}^{d})]\\
& +\lambda(N-i-1)p_{u(i+1,\bar{i}^{d})}[\Delta_{k}(i+2,\bar{i}^{d})-\Delta_{k}(i+1,\bar{i}^{d})]\\
\geq & 2\kappa + \alpha[1-2(\lambda+\mu)-\lambda(N-i)(p_{u(i-1,\bar{i}^{d})}-p_{u(i+1,\bar{i}^{d})})]\epsilon_{k}^{l}\\
\geq & 2\kappa + \alpha[1-2(\lambda+\mu)-\lambda(N-N_{1})]\epsilon_{k}^{l}
\end{array}
\end{equation}
The second inequality holds since the following four terms are greater or equal than $\epsilon_{k}^{l}$ based on the assumption at iteration $k$: $\Delta_{k}(i+1,\bar{i}^{d})-\Delta_{k}(i,\bar{i}^{d})$, $\Delta_{k}(i,\bar{i}^{d})-\Delta_{k}(i-1,\bar{i}^{d})$, $\Delta_{k}(i+1,\bar{i}^{u}+\Delta y)-\Delta_{k}(i,\bar{i}^{u}+\Delta y)$, $\Delta_{k}(i+1,\bar{i}^{d}-\Delta y)-\Delta_{k}(i,\bar{i}^{d}-\Delta y)$. 

Denote $\upsilon=\lambda(N-N_{1})$ and $\epsilon_{k+1}^{l}=2\kappa +\alpha[1-2(\lambda+\mu)-\upsilon]\epsilon_{k}^{l}$. It can be now seen that
\begin{equation}
\Delta_{k+1}(i+1,\bar{i}^{d})-\Delta_{k+1}(i,\bar{i}^{d})\geq \epsilon_{k+1}^{l}
\end{equation}
holds for all $\{i,\bar{i}^{d}\}$ at iteration $k+1$. Similarly we can prove 
\begin{equation}
\Delta_{k+1}(i+1,\bar{i}^{u})-\Delta_{k+1}(i,\bar{i}^{u})\geq \epsilon_{k+1}^{l}.
\end{equation}
By mathematical induction, it is easy to show that
\begin{equation}
\Delta_{k}(i+1,\bar{i}^{d})-\Delta_{k}(i,\bar{i}^{d})\geq \epsilon_{k}^{l}, \Delta_{k}(i+1,\bar{i}^{u})-\Delta_{k}(i,\bar{i}^{u})\geq \epsilon_{k}^{l}
\end{equation}
holds for all $k$ in the infinite series $\epsilon_{k}^{l}$ generated recursively by
\begin{equation}
\epsilon_{k+1}^{l}=2\kappa +\alpha[1-2(\lambda+\mu)-\upsilon]\epsilon_{k}^{l}.
\end{equation}
Since $\epsilon_{0}^{l}=0$, $\epsilon_{k}^{l}$ must converge for $k\rightarrow\infty$. In fact, it converges to $\epsilon^{l}$ with
\begin{equation}
\epsilon^{l} = \lim\limits_{k\rightarrow\infty}\epsilon_{k}^{l} = \frac{2\kappa}{1-\alpha[1-2(\lambda+\mu)-\upsilon]}
\end{equation}
And we can hence conclude that
\begin{displaymath}
\begin{array}{ll}
&\Delta(i+1,\bar{i}^{d})-\Delta(i,\bar{i}^{d})\\
=&\lim\limits_{k\rightarrow\infty}[\Delta_{k}(i+1,\bar{i}^{d})-\Delta_{k}(i,\bar{i}^{d})],\\
\geq & \lim\limits_{k\rightarrow\infty}\epsilon_{k}^{l} = \epsilon^{l}.
\end{array}
\end{displaymath}

(2) Assuming $\Delta(i,\bar{i}^{d})+\epsilon_{k}^{u}\geq \Delta(i+1,\bar{i}^{d})$ holds with $\epsilon_{k}^{u}=0$ for all $\{i,\bar{i}^{d}\}$ and $k=0$, Lemma 2 implies
{\small
\begin{displaymath}
-(\phi(\Delta(i+2,\bar{i}^{d}))-\phi(\Delta(i+1,\bar{i}^{d}))) \leq \alpha p_{u(i,\bar{i}^{d})}[\Delta(i+2,\bar{i}^{d})-\Delta(i+1,\bar{i}^{d})],
\end{displaymath}
\begin{displaymath}
\phi(\Delta(i+1,\bar{i}^{d}))-\phi(\Delta(i,\bar{i}^{d})) \leq -\alpha p_{u(i,\bar{i}^{d})}(\Delta(i+1,\bar{i}^{d}))-\Delta(i,\bar{i}^{d})).
\end{displaymath}
}
Substituting the above two inequalities into (\ref{proposition 2 eq 1}) we get
\begin{displaymath}
\begin{array}{ll}
&\Delta_{k+1}(i+1,\bar{i}^{d})-\Delta_{k+1}(i,\bar{i}^{d})\\
\leq & g(i+1,\bar{i}^{d})-2g(i,\bar{i}^{d})+g(i-1,\bar{i}^{d})\\
&+\alpha\big\{ (1-\gamma_{1}^{d}-\gamma_{2}^{d}-\mu(i+1)-\lambda(N-i+1)p_{u(i,\bar{i}^{d})})\\
& [\Delta_{k}(i+1,\bar{i}^{d})-\Delta_{k}(i,\bar{i}^{d})]\\
& +\mu(i-1)[\Delta_{k}(i,\bar{i}^{d})-\Delta_{k}(i-1,\bar{i}^{d})]\\
& + \gamma_{1}^{d}[\Delta_{k}(i+1,\bar{i}^{u}+\Delta y)-\Delta_{k}(i,\bar{i}^{u}+\Delta y)]\\
& + \gamma_{2}^{d}[\Delta_{k}(i+1,\bar{i}^{d}-\Delta y)-\Delta_{k}(i,\bar{i}^{d}-\Delta y)]\big\}\\
& + \lambda (N-i-1)\alpha p_{u(i,\bar{i}^{d})}\epsilon_{k}^{u}\\
\leq & 2\kappa + \alpha[1-2(\lambda+\mu)]\epsilon_{k}^{u}
\end{array}
\end{displaymath}
The second inequality holds since the following four terms are smaller or equal than $\epsilon_{k}^{u}$ based on our assumption at iteration $k$: $\Delta_{k}(i+1,\bar{i}^{d})-\Delta_{k}(i,\bar{i}^{d})$, $\Delta_{k}(i,\bar{i}^{d})-\Delta_{k}(i-1,\bar{i}^{d})$, $\Delta_{k}(i+1,\bar{i}^{d}+\Delta y)-\Delta_{k}(i,\bar{i}^{d}+\Delta y)$, $\Delta_{k}(i+1,\bar{i}^{d}-\Delta y)-\Delta_{k}(i,\bar{i}^{d}-\Delta y)$. Denoting $\epsilon_{k+1}^{u}=2\kappa + \alpha[1-2(\lambda+\mu)]\epsilon_{k}^{u}$, it can be seen that
\begin{equation}
\Delta_{k+1}(i+1,\bar{i}^{d})-\Delta_{k+1}(i,\bar{i}^{d})\leq\epsilon_{k+1}^{u}
\end{equation}
for all $\{i,\bar{i}^{d}\}$ at iteration $k+1$. Similarly we can prove
\begin{equation}
\Delta_{k+1}(i+1,\bar{i}^{u})-\Delta_{k+1}(i,\bar{i}^{u})\leq\epsilon_{k+1}^{u}
\end{equation}
By mathematical induction, we conclude that
\begin{equation}
\Delta_{k}(i+1,\bar{i}^{d})-\Delta_{k}(i,\bar{i}^{d})\leq \epsilon_{k}^{u}, \Delta_{k}(i+1,\bar{i}^{u})-\Delta_{k}(i,\bar{i}^{u})\leq\epsilon_{k}^{u}
\end{equation}
holds for all $k$ in the infinite series $\epsilon_{k}^{u}$ generated recursively by
\begin{equation}
\epsilon_{k+1}^{u}=2\kappa + \alpha[1-2(\lambda+\mu)]\epsilon_{k}^{u}.
\end{equation}
Since $\epsilon_{0}^{u}=0$, $\epsilon^{u}_{k}$ must converge as $k\rightarrow\infty$. Indeed it converges to $\epsilon^{u}$ with
\begin{equation}
\epsilon^{u} = \lim\limits_{k\rightarrow\infty}\epsilon_{k}^{u} = \frac{2\kappa}{1-\alpha[1-2(\lambda+\mu)]}.
\end{equation}
Hence we have
\begin{displaymath}
\begin{array}{ll}
& \Delta(i+1,\bar{i}^{d})-\Delta(i,\bar{i}^{d})\\
= & \lim\limits_{k\rightarrow\infty} [\Delta_{k}(i+1,\bar{i}^{d})-\Delta_{k}(i,\bar{i}^{d})],\\
\leq & \lim\limits_{k\rightarrow\infty} \epsilon_{k}^{u} = \epsilon^{u}.
\end{array}
\end{displaymath}
And this concludes the proof of Proposition 3. \hfill\(\Box\)
\bibliography{references}{}

\begin{thebibliography}{10}

\bibitem{MakarovLoutan2009}
Yuri~V. Makarov, Clyde Loutan, Jian Ma, and Phillip de~Mello.
\newblock Operational impacts of wind generation on california power systems.
\newblock {\em IEEE Trans. Power Syst.}, 24(2):1039--1050, May 2009.

\bibitem{6344606}
Jian Ma, Shuai Lu, P.V. Etingov, and Y.V. Makarov.
\newblock Evaluating the impact of solar generation on balancing requirements
  in southern nevada system.
\newblock In {\em Power and Energy Society General Meeting, 2012 IEEE}, pages
  1--7, 2012.

\bibitem{5713846}
D.A. Halamay, T.K.A. Brekken, A.~Simmons, and S.~McArthur.
\newblock Reserve requirement impacts of large-scale integration of wind,
  solar, and ocean wave power generation.
\newblock {\em Sustainable Energy, IEEE Transactions on}, 2(3):321--328, 2011.

\bibitem{hammerstrom2007pacific}
Donald~J Hammerstrom, Jerry Brous, David~P Chassin, Gale~R Horst, Robert
  Kajfasz, Preston Michie, Terry~V Oliver, Teresa~A Carlon, Conrad Eustis,
  Olof~M Jarvegren, et~al.
\newblock Pacific northwest gridwise™ testbed demonstration projects; part
  ii. grid friendly™ appliance project.
\newblock Technical report, Pacific Northwest National Laboratory (PNNL),
  Richland, WA (US), 2007.

\bibitem{FERC_2013}
Demand Response and Advanced Metering, FERC staff report, October 2013:
  http://www.pjm.com/$\sim$/media/documents.

\bibitem{LiCaramanis2012}
Ioannis~Ch. Paschalidis, Binbin Li, and Michael~C. Caramanis.
\newblock Demand-side management for regulation service provisioning through
  internal pricing.
\newblock {\em IEEE Trans. Power Systems}, 27(3):1531--1539, August 2012.

\bibitem{BilginCaramanis2012}
Michael~C. Caramanis, Ioannis~Ch. Paschalidis, Christos~G. Cassandras, Enes
  Bilgin, and Elli Ntakou.
\newblock Provision of regulation service reserves by flexible distributed
  loads.
\newblock In {\em the 51th IEEE Conference on Decision and Control}, pages
  3694--3700, 2012.

\bibitem{BowenBaillieul2012}
Bowen Zhang and John Baillieul.
\newblock A packetized direct load control mechanism for demand side
  management.
\newblock In {\em the 51st IEEE Conference on Decision and Control}, pages
  3658--3665, 2012.

\bibitem{Chassin2011-1}
David.~P. Chassin and J.~C. Fuller.
\newblock On the equilibrium dynamics of demand response in thermostatic loads.
\newblock In {\em the 44th Hawaii International Conference on System Sciences},
  2011.

\bibitem{Callaway2011-2}
Stephan Koch, Johanna~L. Mathieu, and Duncan~S. Callaway.
\newblock Modeling and control of aggregated heterogeneous thermostatically
  controlled loads for ancillary services.
\newblock In {\em the Proceedings of the 17th Power Systems Computation
  Conference}, 2011.

\bibitem{6345351}
Wei Zhang, K.~Kalsi, J.~Fuller, M.~Elizondo, and D.~Chassin.
\newblock Aggregate model for heterogeneous thermostatically controlled loads
  with demand response.
\newblock In {\em Power and Energy Society General Meeting, 2012 IEEE}, pages
  1--8, July 2012.

\bibitem{Callaway2009}
Duncan~S Callaway.
\newblock Tapping the energy storage potential in electric loads to deliver
  load following and regulation, with application to wind energy.
\newblock {\em Energy Conversion and Management}, 50(5):1389--1400, 2009.

\bibitem{6760990}
S.~Meyn, P.~Barooah, A~Busic, and J.~Ehren.
\newblock Ancillary service to the grid from deferrable loads: The case for
  intelligent pool pumps in florida.
\newblock In {\em Decision and Control (CDC), 2013 IEEE 52nd Annual Conference
  on}, pages 6946--6953, Dec 2013.

\bibitem{5762580}
M.D. Ilic, Le~Xie, and Jhi-Young Joo.
\newblock Efficient coordination of wind power and price-responsive demand --
  part i: Theoretical foundations.
\newblock {\em Power Systems, IEEE Transactions on}, 26(4):1875--1884, Nov
  2011.

\bibitem{6426122}
M.~Alizadeh, Tsung-Hui Chang, and A~Scaglione.
\newblock Grid integration of distributed renewables through coordinated demand
  response.
\newblock In {\em Decision and Control (CDC), 2012 IEEE 51st Conference on},
  pages 3666--3671, Dec 2012.

\bibitem{conejo2010real}
Antonio~J Conejo, Juan~M Morales, and Luis Baringo.
\newblock Real-time demand response model.
\newblock {\em IEEE Transactions on Smart Grid}, 1(3):236--242, 2010.

\bibitem{6547835}
L.~P. Qian, Y.~J.~A. Zhang, J.~Huang, and Y.~Wu.
\newblock Demand response management via real-time electricity price control in
  smart grids.
\newblock {\em IEEE Journal on Selected Areas in Communications},
  31(7):1268--1280, July 2013.

\bibitem{roozbehani2012volatility}
Mardavij Roozbehani, Munther~A Dahleh, and Sanjoy~K Mitter.
\newblock Volatility of power grids under real-time pricing.
\newblock {\em IEEE Transactions on Power Systems}, 27(4):1926--1940, 2012.

\bibitem{hammerstrom2007pacific1}
DJ~Hammerstrom, R~Ambrosio, J~Brous, TA~Carlon, DP~Chassin, JG~DeSteese,
  RT~Guttromson, GR~Horst, OM~J{\"a}rvegren, R~Kajfasz, et~al.
\newblock Pacific northwest gridwise testbed demonstration projects; part i.
  olympic peninsula project.
\newblock {\em Part I. Olympic Peninsula Project}, 210, 2007.

\bibitem{7346499}
E.~Bilgin, M.C. Caramanis, I.C. Paschalidis, and C.G. Cassandras.
\newblock Provision of regulation service by smart buildings.
\newblock {\em Smart Grid, IEEE Transactions on}, PP(99):1--11, 2015.

\bibitem{LiCaramanis2011}
Ioannis~Ch. Paschalidis, Binbin Li, and Michael~C. Caramanis.
\newblock A market-based mechanism for providing demand-side regulation service
  reserves.
\newblock In {\em the 50th IEEE Conference on Decision and Control and European
  Control Conference}, pages 21--26, December 2011.

\bibitem{Kelly1997}
Frank~P Kelly.
\newblock Charging and rate control for elastic traffic.
\newblock {\em European Transactions on Telecommunications}, 8(1):33--37, 1997.

\bibitem{Kelly1998rate}
Frank~P Kelly, Aman~K Maulloo, and David~KH Tan.
\newblock Rate control for communication networks: shadow prices, proportional
  fairness and stability.
\newblock {\em Journal of the Operational Research society}, 49(3):237--252,
  1998.

\bibitem{PaschalidisTsitsiklis2000}
Ioannis~Ch. Paschalidis and John~N. Tsitsiklis.
\newblock Congestion-dependent pricing of network services.
\newblock {\em IEEE/ACM Trans. Networking}, 8(2):171--184, April 2000.

\bibitem{MutluAlanyaliStarobinskl2000}
Huseyin Mutlu, Murat Alanyali, , and David Starobinski.
\newblock Spot pricing of secondary spectrum access in wireless cellular
  networks.
\newblock {\em IEEE/ACM Trans. Networking}, 17(6):1794--1804, December 2009.

\bibitem{RatepayerRovolt}
Ratepayer revolt in Pacific Northwest: http://news.google.com/
  newspapers?nid=1908\&dat=19820303\&id=j0IrAAAAIBAJ\&sjid=TdQEA
  AAAIBAJ\&pg=3606,5327555.

\bibitem{Bertsekas2007}
Dimitri~P. Bertsekas.
\newblock {\em Dynamic Programming and Optimal Control}.
\newblock Athena Scientific, 2007.

\bibitem{PJMManual2012}
PJM Balancing Operations:
  http://pjm.com/markets-and-operations/ancillary-servicess.

\bibitem{gelenbe2012energy}
Erol Gelenbe.
\newblock Energy packet networks: adaptive energy management for the cloud.
\newblock In {\em Proceedings of the 2nd International Workshop on Cloud
  Computing Platforms}, page~1. ACM, 2012.

\bibitem{Callaway2012}
Johanna~L. Mathieu and Duncan~S. Callaway.
\newblock State estimation and control of heterogeneous thermostatically
  controlled loads for load following.
\newblock In {\em the 45th Hawaii International Conference on System Sciences},
  pages 2002 -- 2011, 2012.

\bibitem{PJMSignal2012}
PJM market and operations:
  http://pjm.com/markets-and-operations/ancillary-services.

\bibitem{BowenBaillieul2013}
Bowen Zhang and John Bailliuel.
\newblock A two level feedback system design to provide regulation reserve.
\newblock In {\em the 52nd IEEE Conference on Decision and Control}, pages
  4322--4328, 2013.

\end{thebibliography}
\bibliographystyle{unsrt}

%

\begin{IEEEbiography}[{\includegraphics[width=1in,height=1.25in,clip,keepaspectratio]{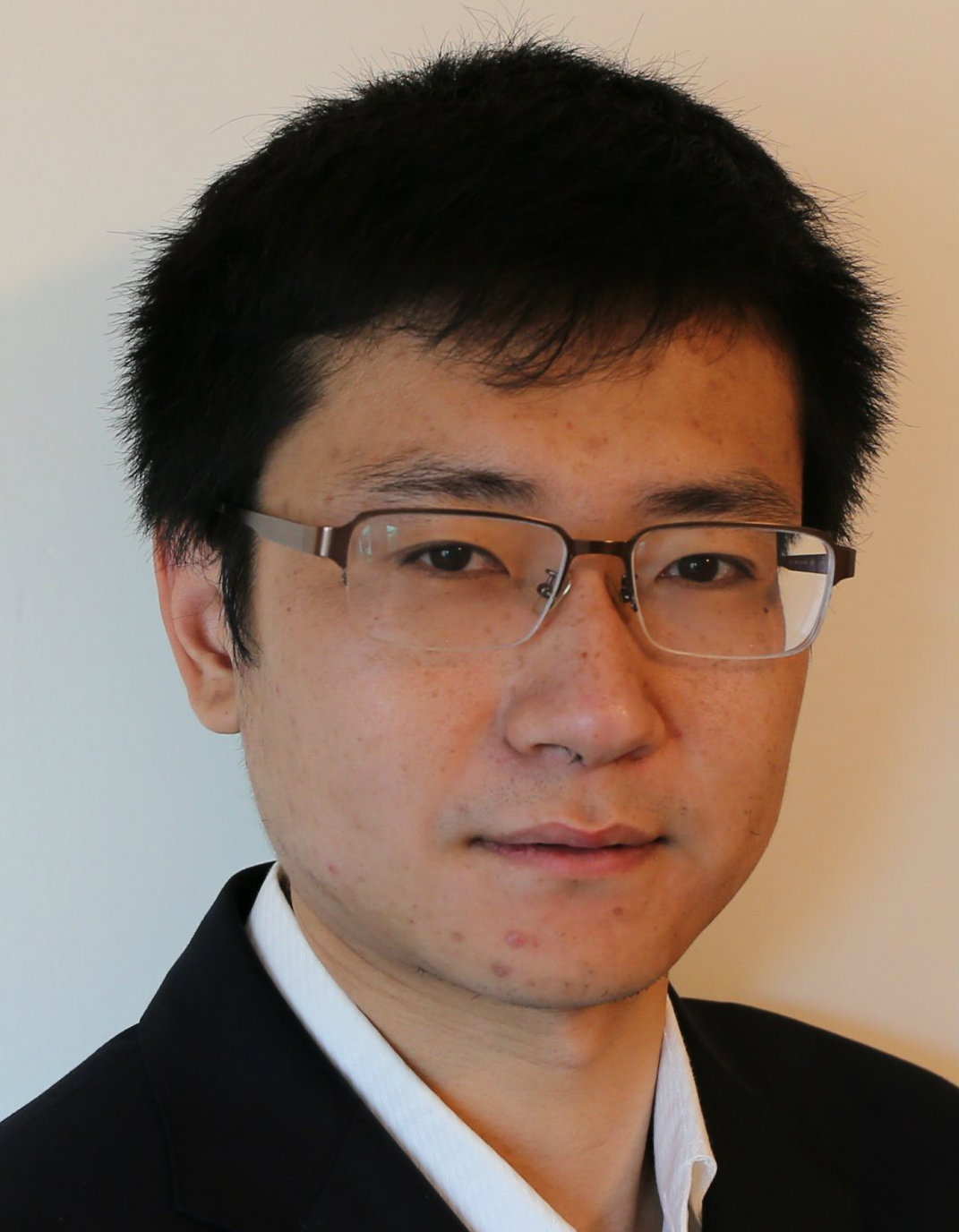}}]{Bowen Zhang}
(M'15) received the B.E. degree in control science and engineering 
from Zhejiang University, Hangzhou, China, in 2010 
and the M.S. and Ph.D degree in Division of Systems Engineering from 
Boston University, Boston, MA, in 2013 and 2015, respectively.
He is currently an Operations Research Analyst at Cambridge Energy Solutions in Cambridge, MA and a visiting scholar to the Department of Mechanical Engineering at Boston University in Boston, MA.

His research focuses on control and optimization 
with applications to electric power system, demand side management, and electricity markets.
\end{IEEEbiography}
%
\begin{IEEEbiography}[{\includegraphics[width=1in,height=1.25in,clip,keepaspectratio]{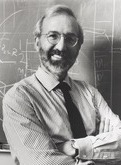}}]{Michael C. Caramanis}
(M'81-SM'13) received the B.S. degree in chemical engineering from Stanford University, Stanford, CA, in 1971 and the M.S. and Ph.D. degrees in engineering with minors in microeconomics and economic development from Harvard University, Cambridge, MA, in 1972 and 1976, respectively.

He is a Professor at Boston University, Boston, MA, with appointments in the Department of Mechanical Engineering and the Division of Systems Engineering. He served at the Greek National Energy council (1976-1979), the MIT Energy Laboratory (1979-1982), and was chair of the Greek Regulatory Authority for Energy (2005-2008). His research focuses on complex stochastic dynamic systems with applications in manufacturing systems and dynamic power markets.
\end{IEEEbiography}


\begin{IEEEbiography}[{\includegraphics[width=1in,height=1.25in,clip,keepaspectratio]{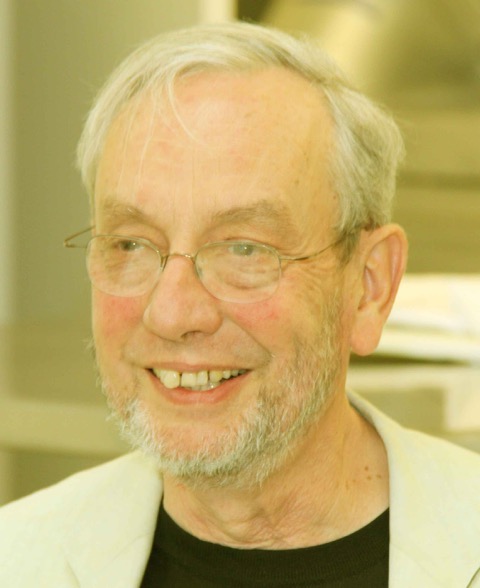}}]{John Baillieul}
(M'83-SM'89-F'93) performs research that deals with mathematical system theory in all of its manifestations and applications. His early
work dealt with connections between optimal control
theory and what came to be called sub-Riemannian
geometry. Other early work treated the control
of nonlinear systems modeled by homogeneous
polynomial differential equations. Such systems
describe, for example, the controlled dynamics of a
rigid body. His main controllability theorem applied
the concept of finiteness embodied in the Hilbert
basis theorem to develop a controllability condition that could be verified by checking the rank of an explicit finite dimensional operator. Over the past decade, he has turned his attention to the study of information based control. Together with K. Li, he has explored source coding of feedback signals that are
designed to provide optimally robust performance in the face of time-varying feedback channel capacity. He has also collaborated with W. S. Wong on research aimed at understanding information complexity in terms of the energy required for two (or more) agents to communicate through the dynamics of a control system. Dr. Baillieul is a Fellow of IFAC and a Fellow of SIAM.
\end{IEEEbiography}




\end{document}